\documentclass[preprint,showpacs,preprintnumbers,amsmath,amssymb]{revtex4}

\usepackage{graphicx}% Include figure files
\usepackage{dcolumn}% Align table columns on decimal point
\usepackage{bm}% bold math
\usepackage{epstopdf}% pdflatex command active even for eps figures
\usepackage{hyperref}% Creates hyperlinks

\begin{document}

\title{Thermodynamic description of the Ising antiferromagnet on triangular lattice with selective dilution by  modified Pair Approximation method}% Force line breaks with \\

\author{T. Balcerzak}
\email{t\_balcerzak@uni.lodz.pl}
\author{K. Sza{\l}owski}
\email{kszalowski@uni.lodz.pl}
\affiliation{%
Department of Solid State Physics, Faculty of Physics and Applied Informatics,\\
University of \L\'{o}d\'{z}, ulica Pomorska 149/153, 90-236 \L\'{o}d\'{z}, Poland
}%

\author{M. Ja\v{s}\v{c}ur}
\author{M. \v{Z}ukovi\v{c}}
\author{A. Bob\'{a}k}
\author{M. Borovsk\'{y}}
\affiliation{%
Department of Theoretical Physics and Astrophysics, Faculty of Science,\\
P. J. \v{S}af\'arik University, Park Angelinum 9, 041 54 Ko\v{s}ice, Slovak Republic
}%

\date{\today}% It is always \today, today,

\begin{abstract}
The Pair Approximation method is modified in order to describe the systems with geometrical frustration. The Ising antiferromagnet on triangular lattice with selective dilution (Kaya-Berker model) is considered and a self-consistent thermodynamic description of this model is obtained. For this purpose, the Gibbs free-energy as a function of temperature, concentration of magnetic atoms on the selected sublattice, and external magnetic field is derived. In particular, the phase diagram is constructed and a comparison of different methods is presented. The thermodynamic quantities are discussed in the context of their physical validity and the improvement in the description introduced by modified method is emphasized.
\end{abstract}

\maketitle

%section 1
\section{Introduction}

Although frustrated spin systems have been studied in literature for over six decades, they still present a challenging problem for theorists 
\cite{Wannier, Blackman,Kawamura,Chowdhury,Fisher,Diep1,Netz,Pelizzola,Kaya,Garcia,Moessner,Galam,Wu,Robinson1,Diep2,Nakatsuji,Ye,Itou,Andrews,Kalz,Lacroix}. During recent years the interest in theoretical studies of such systems is still increasing \cite{Balents,Hartler,Zukovic1,Farnell,Albuquerque,Hauke,Robinson2,Mezzacapo,Fishman,Zukovic2,Shirata,Rojas,Dublenych,Zukovic3,Iglovikov,Maryasin,Nakayama,Melchert,Kulagin,Yokota}. This interest mainly concerns the low-dimensional frustrated magnets which exhibit an intriquing interplay between order and disorder and can reveal existence of new magnetic phases. In turn, the theoretical efforts stimulate search for the experimental realization of such systems. For instance, in response to theoretical demand, frustrated triangular lattices have been synthetized in some layered magnets \cite{Collins,Nakatsuji,Nakatsuji2,Myoung,Cava,Itou,Shirata,Nakayama}.\\

The theoretical studies of low-dimensional frustrated systems have included such structures as: triangular \cite{Wannier,Blackman,Diep1,Pelizzola,Kaya,Moessner,Robinson1,Ye,Zukovic1,Hauke,Zukovic2,Fishman,Zukovic3,Maryasin,Kulagin,Yokota}, honeycomb \cite{Andrews,Farnell,Albuquerque,Mezzacapo}, square \cite{Wu,Kalz,Iglovikov}, triangular-square \cite{Kawamura}, kagom\'{e} \cite{Garcia,Moessner,Loh}, pyrochlore \cite{Garcia}, penthagonal \cite{Rojas}, Shastry-Sutherland lattice \cite{Dublenych} and other \cite{Strecka}. Apart from several exact results \cite{Wannier,Diep1,Wu,Rojas,Dublenych,Melchert} most of them have been obtained by the approximate methods, for example: constant coupling method \cite{Garcia}, Green function technique \cite{Hartler}, spin waves approach \cite{Hauke}, cluster theory \cite{Pelizzola,Farnell,Yokota}, Hard-Spin Mean-Field (HSMF) method \cite{Netz, Kaya}, Effective Field Theory (EFT)\cite{Zukovic1,Zukovic2}, and Monte Carlo (MC) simulations \cite{Kawamura,Robinson1,Andrews,Kalz,
Zukovic3,Iglovikov,Maryasin,Kulagin, Zhang,Jacobsen}.\\

The classical frustrated spin system has been antiferromagnetic Ising model with spin $S=1/2$ on triangular lattice with nearest-neighbour (NN) interactions. This model has been solved exactly by Wannier \cite{Wannier}, who has shown that no long-range ordering exists there at any temperature $T>0$. For $T=0$ the long-range correlation function has been studied in Ref.~\cite{Stephenson} showing its algebraic decay with the distance as $\propto 1/\sqrt{r}$.
On the other hand, the triangular lattice can also be understood as the planar, hexagonal centered lattice \cite{Diep1}. When the central atoms from each hexagon are removed we obtain pure honeycomb lattice. In turn, the honeycomb lattice with antiferromagnetic NN interactions has no frustration and reveals antiferromagnetic ordering.\\

\begin{figure}
\includegraphics[scale=1.0]{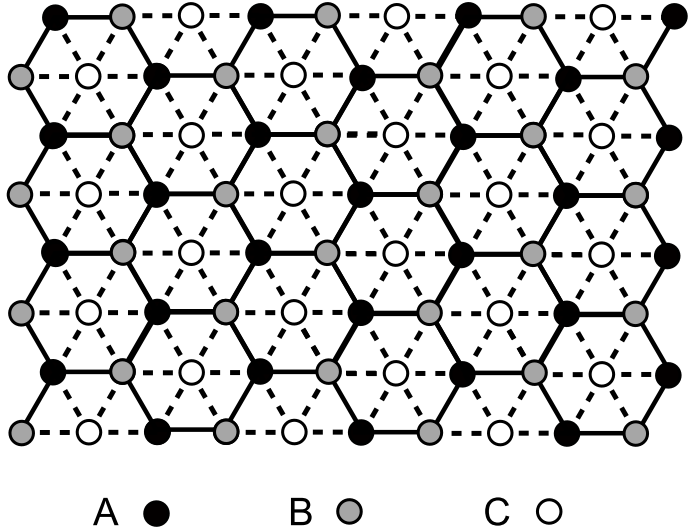}
\caption{A schematic presentation of the triangular lattice consisting of three sublattices $A$, $B$ and $C$. Sublattice $C$ is randomly diluted with concentration $p$.}
\label{fig:fig1}
\end{figure}

According to above observation, Kaya and Berker \cite{Kaya} proposed a model in which the atoms in centres of hexagons are randomly diluted. Such a model presents an intermediate situation between fully frustrated (disordered) triangular lattice and unfrustrated honeycomb lattice with antiferromagnetic order. Namely, in the Kaya-Berker (K-B) model the system can be decomposed into three interpenetrating lattices $A$, $B$, and $C$. The situation is schematically presented in Fig.~\ref{fig:fig1}.
We assume that sublattice $C$ is randomly diluted with concentration $0\leq p \leq1$. In particular, for $p=0$, we obtain pure hexagonal lattice composed of $A$ and $B$ atoms only, whereas for $p=1$, when all sites $C$ are occupied by  magnetic atoms, the system presents an ideal triangular lattice.  Thus, occurrence of selective dilution on $C$-sublattice presents an interesting situation, where the magnetic ordering emerges when $p$-parameter decreases. It is worth-noticing that, when the concentration $p$ is large enough ($p\ge1/2$), a decrease of $p$ means increase of structural disorder, so this phenomenon can be regarded as an structural analogue of "order by disorder" effect.\\
One of the first results for K-B model, which has originally been desribed in the frame of HSMF method \cite{Kaya}, was prediction of the critical concentration $p_c$ for the diluted lattice, below which the system developes long-range ordering. In the first approximation this concentration amounts to $p_c=0.958$, whereas in the further approximation $p_c=0.875$. The last result has recently been confirmed by EFT calculations \cite{Zukovic2}, in astonishing accuracy of three digits. In order to explain this agreement, it one can show on the basis of Ref.~\cite{Balcerzak1} that EFT (which has been introduced some time ago by Honmura and Kaneyoshi \cite{Honmura}) is formally equivalent to HSMF method in its futher approximation. On the other hand, EFT is also equivalent to the I-st Matsudaira approximation \cite{Matsudaira,Balcerzak2}. It has also been shown that HSMF is equivalent to an improved mean-field theory \cite{Banavar,Parisi} which is nothing more but EFT. So that, the coincidence of the results for 
$p_c$, yielded by HSMF and EFT, becomes  
rather obvious consequence. On the other hand, MC studies of K-B model predict, at least in the region of $0\leq p \leq 0.95$, that the antiferromagnetic ordering exists on $A$ and $B$ sublattices, and for $p=0.95$ the critical temperature is still high (about 40\% of its maximal value for $p=0$) \cite{Robinson1}.\\

In the view of above results, the value of $p_c$, and even its existence, remains unsettled. It is worth noticing that existence of $p_c<1$ would imply existence of the interval $p_c<p<1$ in which no ordering takes place at $T=0$. On the other hand, for $p=1$, i.e., for pure triangular lattice it follows from Wannier paper \cite{Wannier} that the system may order at $T=0$ with no costs of energy. This conclusion has also been confirmed in Ref. \cite{Galam}. The ordered state at $T=0$ corresponds to the following sublattice magnetizations: $m_A=1/2$, $m_B=-1/2$, and $m_C=0$, where $A$, $B$, and $C$ are arbitrarily chosen sublattices. In the context of these results, the existence of the gap for $p_c<p<1$, where no ordering takes place, would be difficult to explain. This issue motivated us for studies, by means of another method, whether $p_c<1$ in the K-B model exists, or not.\\

We will apply the Pair Approximation (PA) method in the frame of the Cluster Variational Approach. The method is based on the cumulant expansion for the entropy \cite{Bukman} when the second-order cumulants are taken into account and higher-order cumulants are neglected. This approach has already been applied for the low-dimensional Ising \cite{Balcerzak3} and Heisenberg \cite{Balcerzak4} systems, including structural disorder \cite{Balcerzak5}. The advantage of the PA method over the Molecular Field Approximation (MFA) has been discussed there.
It is worth noticing that, contrary to MFA, PA method takes into account nearest-neighbour spin-pair correlation functions which incorporate important fluctuations.
Recently, the method has also been applied for the ferromagnetic analogue of K-B model, without frustration \cite{Balcerzak6}. However, for the frustrated system considered here the method should be adopted with some necessary modification, which will be explained in the theoretical section.\\

The modified PA method proposed here yields the Gibbs free energy of the system, which is a function of temperature, external field and number of particles (spins). Next, from the expression for the Gibbs energy all thermodynamic quantities can be derived. Thus, the modified PA method gives possibility of complete thermodynamic description of frustrated system in an approximate, but fully self-consistent way.\\

The paper is organized as follows: In the theoretical part a foundation of the PA method will be outlined and its application for frustrated systems will be explained in detail. In further section the numerical calculations will be presented in the figures and discussed. The results concern all basic thermodynamic properties which are obtained from minimization of the Gibbs energy. In particular, the phase diagram and the existence of critical concentration $p_c$ will be discussed in the context of other methods. In the last section a summary of the results will be presented and some final conclusions will be drawn.\\

%section 2
\section{Theoretical model}

\subsection{General formulation}

We consider the Ising model with spins $S_{i_\alpha}=\pm1/2$ arranged on the triangular lattice with antiferromagnetic NN interactions. The $i$-th lattice site belongs to the sublattice $\alpha=A,B,C$, and the random dilution of the selected sublattice ($C$) is assumed. 
The Hamiltonian can be presented as follows:
\begin{equation}
{\mathcal H}=-J\sum_{i_A,j_B}S_{i_A}S_{j_B}-J\sum_{j_B,k_C}S_{j_B}S_{k_C}\xi_{k_C}-J\sum_{i_A,k_C}S_{i_A}S_{k_C}\xi_{k_C}-h\left(\sum_{i_A}S_{i_A}+\sum_{j_B}S_{j_B}+\sum_{k_C}S_{k_C}\xi_{k_C}\right)
\label{eq:eq1}
\end{equation}
where $J\le0$ is the NN antiferromagnetic exchange interaction, $h$ stands for the external field, and $\xi_{k_C}=0,1$ is the random occupation operator.
The configurational mean value of $\xi_{k_C}$, $\left<\xi_{k_C}\right>_r=p$, presents a fraction of occupied sites (concentration of magnetic atoms) on $C$-sublattice.\\

In general, the Gibbs energy $G$ can be presented as:
\begin{equation}
G=\left<{\mathcal H}\right> - ST
\label{eq:eq2}
\end{equation}
where $\left<{\mathcal H}\right>$ is the enthalpy and $S$ presents the entropy of the system. The enthalpy (which is the averaged Hamiltonian containing the external field term) is of the form:
\begin{equation}
\left<{\mathcal H}\right>=-NJ\left(c_{A B}+pc_{B C}+pc_{A C}\right)-\frac{1}{3}Nh\left(m_{A}+m_{B}+pm_{C}\right)
\label{eq:eq3}
\end{equation}
In Eq.~(\ref{eq:eq3}) $N$ denotes the total number of lattice sites in the triangular lattice (which is equal to number of NN lattice site pairs in two sublattices).  The thermal mean values are written in shorten notation as $c_{A B}=\left<S_{i_A}S_{j_B}\right>$, $c_{B C}=\left<S_{j_B}S_{k_C}\right>$,  $c_{A C}=\left<S_{i_A}S_{k_C}\right>$ and denote three NN correlation functions (for occupied lattice sites), whereas $m_{\alpha}=\left<S_{i_\alpha}\right>$ ($\alpha=A,B,C$) denote three sublattice magnetizations per  (occupied) lattice site. Assuming that occupation operators are independent on the Ising spins, the equation (\ref{eq:eq3}) is exact for the model in question.\\

As a general approach, the entropy can be expressed in a series of cumulants \cite{Bukman}. In the PA method only first- and second-order cumulants are taken into account. Thus, the entropy can be approximately presented as follows:
\begin{equation}
S=\frac{N}{3}\left(\sigma_{A}+\sigma_{B}+p\sigma_{C}\right)+N\left(\sigma_{AB}-\sigma_{A}-\sigma_{B}\right)+Np\left(\sigma_{BC}-\sigma_{B}-\sigma_{C}\right)+Np\left(\sigma_{AC}-\sigma_{A}-\sigma_{C}\right)
\label{eq:eq4}
\end{equation}
where $\sigma_{\alpha}$ ($\alpha=A,B,C$) are the single-site entropies and $\sigma_{\alpha \beta}$ ($\alpha \ne \beta=A,B,C$) are the entropies of NN pairs. Expression (\ref{eq:eq4}) for the entropy can be re-written in a more convenient form:
\begin{equation}
S=N\left[\sigma_{AB}+p\left(\sigma_{BC}+\sigma_{AC}\right)-\left(\frac{2}{3}+p\right)\left(\sigma_{A}+\sigma_{B}\right)-\frac{5}{3}p\sigma_{C}\right]
\label{eq:eq5}
\end{equation}
The single site and pair entropies can be found from their definitions:
\begin{equation}
\sigma_{\alpha}=-k_{\rm B}{\rm Tr}_{i_{\alpha}} \left( \rho_{i_{\alpha}} {\rm ln} \rho_{i_{\alpha}}\right)
\label{eq:eq6}
\end{equation}
and
\begin{equation}
\sigma_{\alpha \beta}=-k_{\rm B}{\rm Tr}_{i_{\alpha}j_{\beta}} \left( \rho_{i_{\alpha}j_{\beta}} {\rm ln} \rho_{i_{\alpha}j_{\beta}}\right)
\label{eq:eq7}
\end{equation}
where $\rho_{i_{\alpha}}$ and $\rho_{i_{\alpha}j_{\beta}} $ are the single-site and pair density matrices, respectively. For spins 1/2 these matrices are given by the expressions:
\begin{equation}
\rho_{i_{\alpha}}= \frac{1}{2}+2m_{\alpha}S_{i_{\alpha}}
\label{eq:eq8}
\end{equation}
($\alpha=A,B,C$; $S_{i_{\alpha}}=\pm \frac{1}{2}$), and 
\begin{equation}
\rho_{i_{\alpha}j_{\beta}}= \frac{1}{4}+m_{\alpha}S_{i_{\alpha}}+m_{\beta}S_{j_{\beta}}+4c_{\alpha \beta}S_{i_{\alpha}}S_{j_{\beta}}
\label{eq:eq9}
\end{equation}
($\alpha \ne \beta=A,B,C$),
respectively. Let us remark, that  decoupling of correlations, $c_{\alpha \beta}\approx m_{\alpha}m_{\beta}$, is equivalent to factorization of the pair matrices, $\rho_{i_{\alpha}j_{\beta}}\approx \rho_{i_{\alpha}}\rho_{j_{\beta}}$, and leads to the approximation $\sigma_{\alpha \beta}\approx \sigma_{\alpha}+\sigma_{\beta}$. This is equivalent to the Molecular Field Approximation method, in which the expression (\ref{eq:eq4}) for the entropy contains only the first, additive term \cite{Balcerzak3}. As discussed in Ref.~\cite{Balcerzak3}, the single-site density matrices are normalized:
\begin{equation}
{\rm Tr}_{i_{\alpha}}\rho_{i_{\alpha}}= 1
\label{eq:eq10}
\end{equation}
and the pair density matrices can be reduced:
\begin{equation}
{\rm Tr}_{j_{\beta}}\rho_{i_{\alpha}j_{\beta}}= \rho_{i_{\alpha}}
\label{eq:eq11}
\end{equation}
The matrices given by Eqs.~(\ref{eq:eq8}) and (\ref{eq:eq9}) satisfy the relationships for the thermodynamic mean values:
\begin{equation}
m_{\alpha}=\left<S_{i_\alpha}\right>={\rm Tr}_{i_\alpha}\left(S_{i_\alpha}\rho_{i_{\alpha}} \right)
\label{eq:eq12}
\end{equation}
and
\begin{equation}
c_{\alpha \beta}=\left<S_{i_\alpha}S_{j_\beta}\right>={\rm Tr}_{i_\alpha j_\beta}\left(S_{i_\alpha}S_{j_\beta}\rho_{i_{\alpha}j_{\beta}} \right)
\label{eq:eq13}
\end{equation}
With the help of the density matrices (\ref{eq:eq8}) and (\ref{eq:eq9}) the single-site (\ref{eq:eq6}) and pair (\ref{eq:eq7}) entropies can be expressed as:
\begin{equation}
\sigma_{\alpha}=-k_{\rm B} \left(\frac{1}{2}+m_{\alpha}\right){\rm ln}\left(\frac{1}{2}+m_{\alpha} \right)-k_{\rm B} \left(\frac{1}{2}-m_{\alpha}\right){\rm ln}\left(\frac{1}{2}-m_{\alpha} \right)
\label{eq:eq14}
\end{equation}
($\alpha=A,B,C$), and
\begin{equation}
\sigma_{\alpha \beta}=-k_{\rm B}\rho^{++}_{\alpha \beta}{\rm ln}\rho^{++}_{\alpha \beta}-k_{\rm B}\rho^{+-}_{\alpha \beta}{\rm ln}\rho^{+-}_{\alpha \beta}-k_{\rm B}\rho^{-+}_{\alpha \beta}{\rm ln}\rho^{-+}_{\alpha \beta}-k_{\rm B}\rho^{--}_{\alpha \beta}{\rm ln}\rho^{--}_{\alpha \beta}
\label{eq:eq15}
\end{equation}
($\alpha \ne \beta=A,B,C$), respectively. In Eq.~(\ref{eq:eq15}) we introduced the abbreviate notation:
\begin{eqnarray}
\rho^{++}_{\alpha \beta}&=&\frac{1}{4}+\frac{1}{2}m_{\alpha}+\frac{1}{2}m_{\beta}+c_{\alpha \beta}\nonumber\\
\rho^{+-}_{\alpha \beta}&=&\frac{1}{4}+\frac{1}{2}m_{\alpha}-\frac{1}{2}m_{\beta}-c_{\alpha \beta}\nonumber\\
\rho^{-+}_{\alpha \beta}&=&\frac{1}{4}-\frac{1}{2}m_{\alpha}+\frac{1}{2}m_{\beta}-c_{\alpha \beta}\nonumber\\
\rho^{--}_{\alpha \beta}&=&\frac{1}{4}-\frac{1}{2}m_{\alpha}-\frac{1}{2}m_{\beta}+c_{\alpha \beta}
\label{eq:eq16}
\end{eqnarray}
Taking into account the above formulas and Eq.~(\ref{eq:eq2}), the Gibbs energy per  lattice site, expressed in $|J|$ units, can be written in the final form:
\begin{equation}
\frac{G}{N|J|}=c_{A B}+p\left(c_{A C}+c_{B C}\right)-\frac{1}{3}\,\frac{h}{|J|}\left(m_{A}+m_{B}+pm_{C}\right)-\frac{k_{\rm B}T}{|J|}\,\frac{S}{Nk_{\rm B}}
\label{eq:eq17}
\end{equation}
where the dimensionless entropy per  lattice site is in the form of:
\begin{eqnarray}
\frac{S}{Nk_{\rm B}}=&-&\rho^{++}_{A B}{\rm ln}\rho^{++}_{A B}-\rho^{+-}_{A B}{\rm ln}\rho^{+-}_{A B}-\rho^{-+}_{A B}{\rm ln}\rho^{-+}_{A B}-\rho^{--}_{A B}{\rm ln}\rho^{--}_{A B}\nonumber\\
&-&p\,\left[\rho^{++}_{A C}{\rm ln}\rho^{++}_{A C}+\rho^{+-}_{A C}{\rm ln}\rho^{+-}_{A C}+\rho^{-+}_{A C}{\rm ln}\rho^{-+}_{A C}+\rho^{--}_{A C}{\rm ln}\rho^{--}_{A C}  \right]\nonumber\\
&-&p\,\left[\rho^{++}_{B C}{\rm ln}\rho^{++}_{B C}+\rho^{+-}_{B C}{\rm ln}\rho^{+-}_{B C}+\rho^{-+}_{B C}{\rm ln}\rho^{-+}_{B C}+\rho^{--}_{B C}{\rm ln}\rho^{--}_{B C}  \right]\nonumber\\
&+&\left(\frac{2}{3}+p\right)\left[\left(\frac{1}{2}+m_{A}\right){\rm ln}\left(\frac{1}{2}+m_{A} \right)+\left(\frac{1}{2}-m_{A}\right){\rm ln}\left(\frac{1}{2}-m_{A} \right)\right]\nonumber\\
&+&\left(\frac{2}{3}+p\right)\left[\left(\frac{1}{2}+m_{B}\right){\rm ln}\left(\frac{1}{2}+m_{B} \right)+\left(\frac{1}{2}-m_{B}\right){\rm ln}\left(\frac{1}{2}-m_{B} \right)\right]\nonumber\\
&+&\frac{5}{3}p\,\left[\left(\frac{1}{2}+m_{C}\right){\rm ln}\left(\frac{1}{2}+m_{C} \right)+\left(\frac{1}{2}-m_{C}\right){\rm ln}\left(\frac{1}{2}-m_{C} \right)\right]
\label{eq:eq18}
\end{eqnarray}
and $\rho^{\pm \pm}_{\alpha \beta}$ are given by Eqs.~(\ref{eq:eq16}).

\subsection{Modification of the PA method for the system with geometrical frustration}

From the general formulation, the Gibbs energy given by Eq.~(\ref{eq:eq17}) is a function of 6 variational parameters: $m_A$, $m_B$, $m_C$, $c_{A B}$, $c_{A C}$ and $c_{B C}$. In the conventional approach within PA, the Gibbs energy in equilibrium corresponds to the minimum with respect to all these parametes, which are treated equally. However, in the case of geometrical frustration such treatment leads to wrong (unphysical) results since the correlations $c_{\alpha \beta}$ are not fully independent parameters \cite {Pelizzola}.
As a consequence, the ground-state energy is incorrect and the entropy is negative in the low-temperature region. 
In order to improve on the method, we propose its modification for the Gibbs energy calculation. The modified method is based on the assumption that the correlations involving frustrated spins should be partly decoupled. Namely, let us assume that the spin $S_{k_C}$ is frustrated in $k$-site of the selected lattice $C$. Then, $m_A$, $m_B$, $m_C$ and $c_{A B}$ can  further be treated as independent parameters, however, the correlations $c_{A C}$ and $c_{B C}$, which involve 
the frustrated spin $S_{k_C}$, are not  independent on the rest of parameters and should be treated in a more complex way. First of all, let us observe that in a given triangle ($i_A, j_B, k_C$),  the spin $S_{k_C}$ is  frustrated only when the spins $S_{i_A}$ and $S_{j_B}$ take the antiparallel orientation. The probability $x$ of such situation can be estimated as follows:
\begin{equation}
x=\rho^{+-}_{A B}+\rho^{-+}_{A B}=\frac{1}{2}-2c_{A B}
\label{eq:eq19}
\end{equation}
In the rest of states (i.e., when the spins $S_{i_A}$ and $S_{j_B}$ are parallel) the spin $S_{k_C}$  is not frustrated in this triangle. Such (unfrustrated) situation occurs with the probability $1-x$, where
\begin{equation}
1-x=\rho^{++}_{A B}+\rho^{--}_{A B}=\frac{1}{2}+2c_{A B}
\label{eq:eq20}
\end{equation}
Therefore, for the correlations $c_{A C}$ and $c_{B C}$ we propose the following approximation:
\begin{equation}
c_{A C} \approx x\left<S_{i_A}S_{k_C}\right>^{\prime \prime}+\left(1-x \right)\left<S_{i_A}S_{k_C}\right>^{\prime}
\label{eq:eq21}
\end{equation}
and
\begin{equation}
c_{B C} \approx x\left<S_{j_B}S_{k_C}\right>^{\prime \prime}+\left(1-x \right)\left<S_{j_B}S_{k_C}\right>^{\prime},
\label{eq:eq22}
\end{equation}
respectively. For the case when the spins $S_{i_A}$ and $S_{j_B}$  are parallel, both correlations involving spin $S_{k_C}$ must be equal, therefore $\left<S_{i_A}S_{k_C}\right>^{\prime}=\left<S_{j_B}S_{k_C}\right>^{\prime} = c'$, and  $c'$ can be treated as a new variational parameter, in addition to $m_A$, $m_B$, $m_C$ and $c_{A B}$. On the other hand, for the frustrated states of $S_{k_C}$, the correlations $\left<S_{i_A}S_{k_C}\right>^{\prime \prime}$ and $\left<S_{j_B}S_{k_C}\right>^{\prime \prime}$ should be decoupled as follows:
\begin{equation}
\left<S_{i_A}S_{k_C}\right>^{\prime \prime} \approx \left<S_{i_A}\right>^{\prime \prime}\left<S_{k_C}\right> = \left<S_{i_A}\right>^{\prime \prime} m_C
\label{eq:eq23}
\end{equation}
and
\begin{equation}
\left<S_{j_B}S_{k_C}\right>^{\prime \prime} \approx \left<S_{j_B}\right>^{\prime \prime}\left<S_{k_C}\right> = \left<S_{j_B}\right>^{\prime \prime} m_C,
\label{eq:eq24}
\end{equation}
where $\left<S_{i_A}\right>^{\prime \prime}$ (and $\left<S_{j_B}\right>^{\prime \prime}$) denote the conditional averages, i.e., the averages when the neighbouring spins $S_{j_B}$ and $S_{i_A}$  are antiparallel, respectively. These mean values can be calculated with the help of the two normalized  probabilities: $\rho^{+-}_{A B}/x$ and $\rho^{-+}_{A B}/x$ as follows:
\begin{equation}
 \left<S_{i_A}\right>^{\prime \prime} = \frac{1}{x} \left( \frac{1}{2} \rho^{+-}_{A B} - \frac{1}{2} \rho^{- +}_{A B} \right) = \frac{1}{2x} \left( m_A - m_B \right)
\label{eq:eq25}
\end{equation}
and
\begin{equation}
 \left<S_{j_B}\right>^{\prime \prime} = \frac{1}{x} \left( \frac{1}{2} \rho^{- +}_{A B} - \frac{1}{2} \rho^{+ -}_{A B} \right) = \frac{1}{2x} \left( m_B - m_A \right).
\label{eq:eq26}
\end{equation}
Thus, for the correlations $c_{A C}$ and $c_{B C}$ we obtain the following approximation:
\begin{equation}
 c_{A C} \approx \frac{1}{2} \left( m_A - m_B \right) m_C + \left( \frac{1}{2}+2c_{A B}\right) c'
\label{eq:eq27}
\end{equation}
and
\begin{equation}
 c_{B C} \approx \frac{1}{2} \left( m_B - m_A \right) m_C + \left( \frac{1}{2}+2c_{A B}\right) c'
\label{eq:eq28}
\end{equation}
This approximation contains partial decoupling, but also introduces a new variational parameter $c'$ for the corelations containing unfrustrated states of $S_{k_C}$. Substituting Eqs.~(\ref{eq:eq27}) and  (\ref{eq:eq28}) into the Gibbs energy (\ref{eq:eq17}) (and entropy (\ref{eq:eq18}))  we can describe our frustrated system with no risk of getting unphysical solutions. The equilibrium for the Gibbs energy is obtained for 5 variational parameters only, whose values should be restricted to the following physical ranges: $-1/2 \leq m_{\alpha} \leq 1/2$ ($\alpha = A, B, C$), and $-1/4 \leq (c_{A B}, \, c') \leq 1/4$).\\

\subsection{Thermodynamic properties and the variational equations}

The complete and self-consistent thermodynamic description can be obtained from the basic equation for the Gibbs potential (\ref{eq:eq17}) with the help of Eqs.~(\ref{eq:eq27}) and (\ref{eq:eq28}). Since the Gibbs energy is a function of the external field $h$ and temperature $T$, the first derivatives lead to the results:
\begin{equation}
\frac{1}{N}\, \left(\frac{\partial G}{\partial h}\right)_{T} = - \frac{1}{3} \left( m_A + m_B + pm_C \right) = - m
\label{eq:eq29}
\end{equation}
where $m$ is the averaged magnetization per  lattice site, and
\begin{equation}
\left(\frac{\partial G}{\partial T}\right)_{h} = - S
\label{eq:eq30}
\end{equation}
where $S$ is the entropy (given in the form of Eq.~(\ref{eq:eq18})). It is worth-noticing that formulas (\ref{eq:eq29}) and (\ref{eq:eq30}) are only satisfied together with the necessary extremum conditions:
\begin{equation}
\frac{\partial G}{\partial m_{\alpha}} = 0
\label{eq:eq31}
\end{equation}
($\alpha = A, B, C$),
\begin{equation}
\frac{\partial G}{\partial c_{A B}} = 0
\label{eq:eq32}
\end{equation}
and 
\begin{equation}
\frac{\partial G}{\partial c'} = 0
\label{eq:eq33}
\end{equation}
(provided $|c'| \le 1/4$). 
Eqs.~(\ref{eq:eq31})-(\ref{eq:eq33}) form a set of 5 variational equations from which the variational parameters can be obtained. The detailed form of these equations for $h=0$ is presented in Appendix. When solving such equations it should be controlled whether the solutions fall into the physical ranges $-1/2 \leq m_{\alpha} \leq 1/2$ ($\alpha = A, B, C$), $-1/4 \leq c_{A B}\leq 1/4$) and $-1/4 \leq c' \leq 1/4$). If for a certain parameter this is not the case, we should assume the value of that parameter at the edge of the physical range, where the Gibbs energy reaches its minimum. In such a case, when the variational parameter is constant at the edge, the corresponding variational equation should be ignored.  This objection mainly concerns Eq.~(\ref{eq:eq33})  in the low temperature region, and the consequences are discussed in the next Section in more detail.\\

As far as other thermodynamic properties are concerned, they can be obtained from the second derivatives of the Gibbs energy. For instance, the isothermal susceptibility is given by:
\begin{equation}
\chi_T=N\left(\frac{\partial m}{\partial h}\right)_{T} = -\left(\frac{\partial^2 G}{\partial h^2}\right)_{T}.
\label{eq:eq34}
\end{equation}
In turn, the magnetic contribution to the specific heat at constant field $h$ can be found from the relationship:
\begin{equation}
C_h=T\left(\frac{\partial S}{\partial T}\right)_{h} = -T\left(\frac{\partial^2 G}{\partial T^2}\right)_{h}.
\label{eq:eq35}
\end{equation}
Since the whole theory is self-consistent, the specific heat can also be calculated in the equivalent way:
\begin{equation}
C_h=\left(\frac{\partial \left<{\mathcal H}\right>}{\partial T}\right)_{h}
\label{eq:eq36}
\end{equation}
where $\left<{\mathcal H}\right>$ is the enthalpy given by Eq.~(\ref{eq:eq3}). 
Equivalency of Eqs.~(\ref{eq:eq35}) and (\ref{eq:eq36}) requires that calculations of the entropy must be consistent with calculations of the correlation function.
The numerical results and their detailed analysis will be presented in the next Section.\\

%section 3
\section{Numerical results and discussion}

\begin{figure}
\includegraphics[scale=1.0]{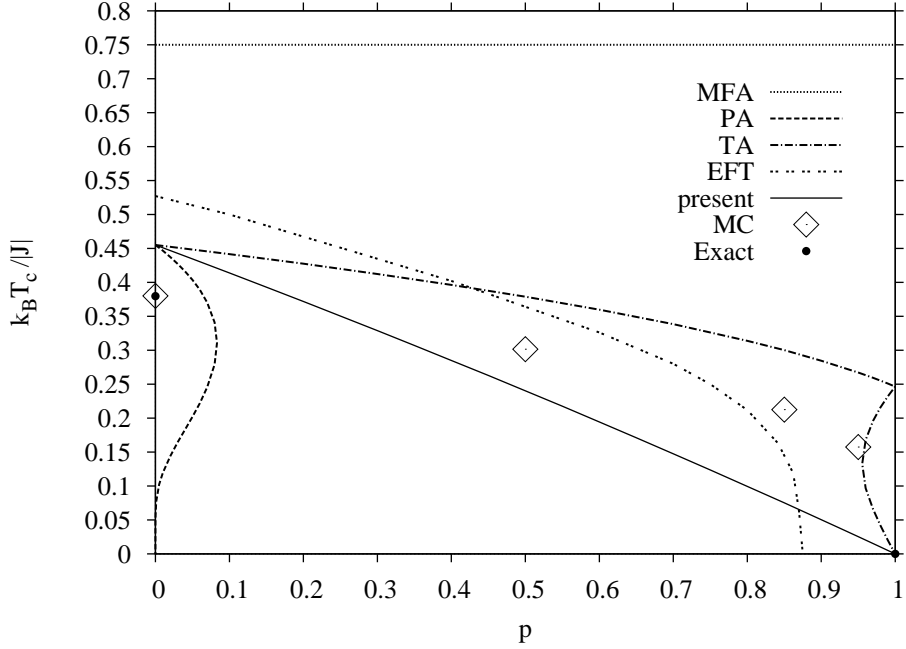}
\caption{Dependence of critical temperature $k_{\rm B} T_{\rm c}/|J|$ on concentration $p$. A comparison of different methods is presented.}
\label{fig:fig2}
\end{figure}
We shall start the numerical analysis from the ground state. At $T=0$ the entropic part in the Gibbs energy is unimportant and only enthalpy (mean value of the Hamiltonian) determines the thermodynamic potential. 
Therefore, the ground state phase diagram can be determined exactly.
By analysis of the enthalpy minimum in 5-dimensional space, in the range of concentration $0 < p < 1$ and $h=0$, we found that the sublattice magnetizations in the ground state take the values: $m_A=1/2$, $m_B=-1/2$ and $m_C=0$ (or, symmetrically, $m_A=-1/2$, $m_B=1/2$ and $m_C=0$). At the same time, the correlation parameters in this regime are: $c_{A B}=-1/4$ and $c'=-1/4$. Absence of the magnetization on $C$-sublattice is due to the fact that the frustrated spins $S_{k_C}$ can take the values $\pm 1/2$ with the same probability. For $h/|J|$ belonging to the range $0 < h/|J| < 1.5$ magnetization in the ground state is given by $m_A=1/2$, $m_B=-1/2$ and $m_C=1/2$ (or $m_A=-1/2$, $m_B=1/2$ and $m_C=1/2$). At $h/|J|=1.5$ the spin-flip transition takes place, and for $1.5 < h/|J| < 3$ the ground state is characterized by $m_A=1/2$, 
$m_
B=1/2$ and $m_
C=-1/2$. The next spin reversal on $C$-sublattice is observed for $h/|J| = 3$ leading to the uniform magnetization $m_A=1/2$, $m_B=1/2$ and $m_C=1/2$ when $h/|J| > 3$.
At $h/|J| = 1.5$ and $h/|J| = 3$ the coexistence of neighbouring phases take place.\\
\begin{figure}
\includegraphics[scale=1.0]{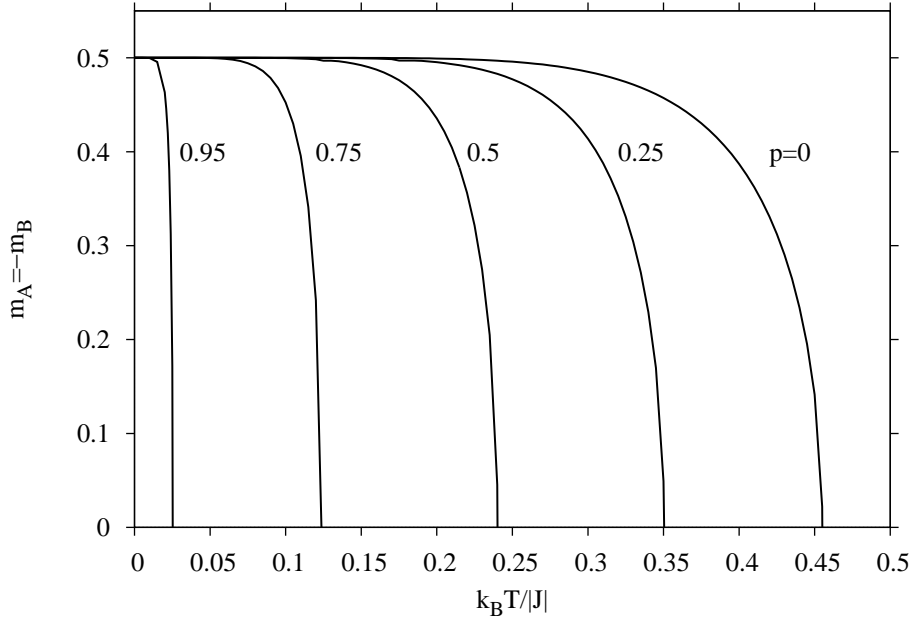}
\caption{Sublattice spontaneous magnetizations $m_A=-m_B$ vs. dimensionless temperature $k_{\rm B} T/|J|$. Different curves correspond to various concentrations $p$.}
\label{fig:fig3}
\end{figure}
For $p=1$ the situation becomes more complex, because for $h=0$ and $T=0$ each triangle consisting of NN spins is 6-fold degenerated \cite{Schick}. From the analysis of enthalpy at this point we found that two distinct kinds of states can coexist with the same energy. One is the ordered state, in which two sublattices have opposite magnetizations, and the magnetization of the third sublattice is equal to zero. This state can be considered as a continuation of the situation which occurs for $0 < p < 1$, and can be characterized by the parameters: $m_A=1/2$, $m_B=-1/2$, $m_C=0$,  $c_{A B}=-1/4$, $c_{A C}=0$ and $c_{B C}=0$. The ordered state  vanishes discontinuously when $T>0$. Another state, which coexists in the ground-state point ($p=1$, $h=0$, $T=0$), is the disordered state. It is characterized by the parameters: $m_A=0$, $m_B=0$, $m_C=0$, $c_{A B}=-1/12$, $c_{A C}=-1/12$ and $c_{B C}=-1/12$. This disordered state extends over non-zero temperatures $T>0$. For $p=1$ and $0< h/|J| < 3$ the ground state is 
ordered, and is characterized by two sublattices oriented in parallel to the external field and one antiparallel. At $h/|J| = 3$ the spin-flip transition takes place and for  $h/|J| > 3$ all three sublattices have magnetizations oriented in parallel with the field.\\
\begin{figure}
\includegraphics[scale=1.0]{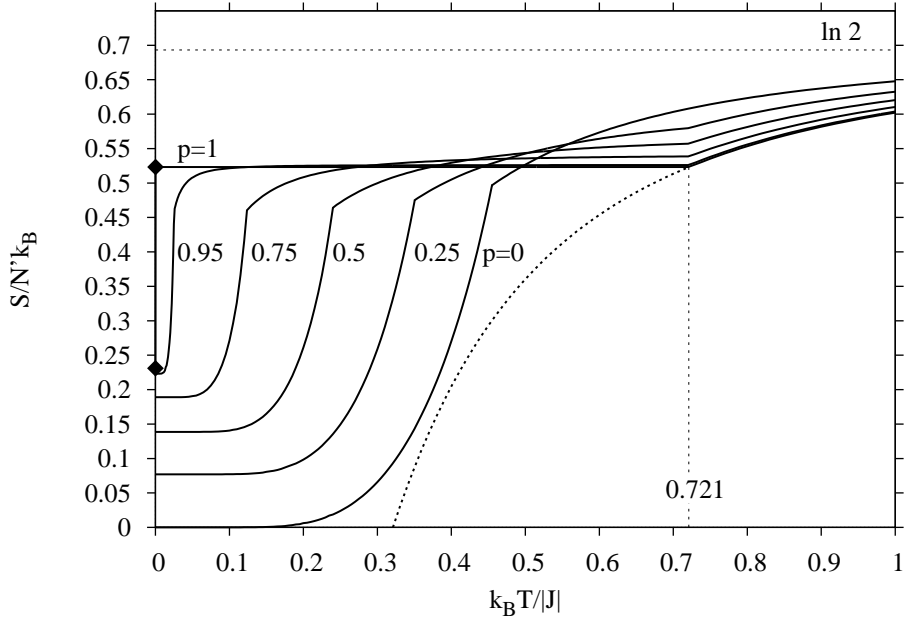}
\caption{Entropy $S/N'k_{\rm B}$ per occupied lattice site, expressed in Boltzmann constant units, vs. dimensionless temperature $k_{\rm B} T/|J|$. Different curves correspond to various concentrations $p$.}
\label{fig:fig4}
\end{figure}
Further, we will concentrate on the numerical calculations of thermodynamic quantities for $h=0$ in the whole concentration range $0 \le p \le 1$ and arbitrary $T$. The most intriguing problem concerns the existence of phase transitions. In Fig.~\ref{fig:fig2} we illustrate the phase transition (N\'eel) temperature vs. concentration $p$. The ordered state presents a continuation of the phase existing in the ground state and is characterized by $m_A=-m_B$ and $m_C=0$. Such solution has also been found by MC simulations \cite{Robinson1}. Various curves and markers in Fig.~\ref{fig:fig2} correspond to different methods. The exact results (marked by bold points) have been found for $p=0$ and $p=1$ as 
$k_{\rm B} T_{\rm c}/|J|=1/\left(2\ln(2+\sqrt{3})\right)\approx 0.3797$  \cite {Wannier2} and   $k_{\rm B} T_{\rm c}/|J|=0$ \cite{Wannier}, respectively. MC results are marked by diamond symbols \cite{Robinson1}. For $p=0$, a good agreement of the MC result with the exact solution can be noted. \\

It should also be mentioned that in case of honeycomb lattice, as well as for other 2D lattices, High Temperature Series Expansion (HTSE) method \cite{Domb} gives the critical temperature which is practically exact. Other approximate methods are not so accurate. For instance, the thermodynamic perturbation theory \cite{Dresselhaus} gives for honeycomb lattice the result $k_{\rm B} T_{\rm c}/|J|=0.43$ (for $S=1/2$) in the 4-th approximation, which is slightly better than Bethe result (0.4551). However, in the 6-th approximation the method developed in Ref.~\cite{Dresselhaus} gives $0.481$, which is worse than the value estimated in the 4-th approximation. For this reason such kind of theory cannot be recommended for honeycomb lattice as a systematic approach. One of recent results for the critical temperature of honeycomb lattice was obtained by correlated cluster mean-field (CCMF) theory \cite{Yamamoto}. The value obtained there was $\approx 0.398$. Also a short overview of other approximate methods can be found in Ref.~\cite{Yamamoto}. However, all these methods have not been applied for Kaya-Berker model with geometrical frustration.\\

It has been known that MC method  is difficult to apply for this model for $0<p<1$ in low-temperature region, where the spins can be frozen and the algorithm becomes trapped in the vicinity of a local free energy minimum \cite{Robinson1}. For $0<p<1$ this difficulty is attributed to glassy behaviour of spins $S_{k_C}$ \cite{Robinson1}. 
It should be mentioned that spin-glass state doesn't occur for $p=1$ (i.e., for pure triangular lattice) \cite{Crest} and MC methods have been successful there for relatively low temperatures \cite{Zhang,Jacobsen}.
However, for $0<p<1$, some analytical methods, which are able to overcome this difficulty, are still desired. The most crude description, which neglects fluctuations, is given by the Molecular Field Approximation (MFA), and is depicted by the horizontal dotted line at $k_{\rm B} T_{\rm c}/|J|=0.75$. It is obvious that such a method cannot be accepted for frustrated systems. Much better description is provided by EFT \cite{Zukovic2} (marked by double-dotted line), which, as discussed in Introduction, is equivalent to HSMF theory in further approximation \cite{Kaya}. In such a case the critical concentration $p_c=0.875$ is predicted, below which the system becomes ordered. In turn, application of the PA method in its conventional formulation (with 6 variational parameters) gives the re-entrant magnetism (dashed curve) and unphysical result, since no ordering is predicted for all $p$. It is worth a mention that for 
ferromagnetic 
systems, without frustration, the PA method gives usually better results that EFT, as for example discussed in Ref. \cite{Balcerzak6}.\\

We are aware that PA method in its usual, unmodified formulation for frustrated systems overestimates NN correlation functions in low temperatures, which leads to improper ground-state energy and negative entropy. Therefore we first tried to develop Triangle Approximation (TA) according to the general Cluster Variational Method (CVM) for the Ising systems \cite {Katsura}. In the case of TA we deal with 12 variational parameters. In some testing calculations for the triangular lattice with ferromagnetic interactions we obtained the Curie temperature $k_{\rm B} T_{\rm c}/|J|=1/\left(2\ln(5/3)\right)\approx$0.9788, which is not far from the exact result $k_{\rm B} T_{\rm c}/J=1/\ln3\approx$0.9102 \cite {Wannier2}. However, for the frustrated model with antiferromagnetic interactions the result of TA method (marked by the dashed-dotted line in Fig.~\ref{fig:fig2}) is still unsatisfactory. In particular, the re-entrant magnetism is found near $p=1$, and for $p=1$ the ordered phase extends up to $k_{\rm B} T_{\rm 
c}/|J|=0.2464$. It is expected that further 
approximations within systematic CVM will improve on the results, however, the number of variational parameters then increases (for instance, there would be 21 parameters in Hexagonal Approximation) and the method developed in Ref. \cite{Katsura} becomes not tractable in practice.\\
\begin{figure}
\includegraphics[scale=1.0]{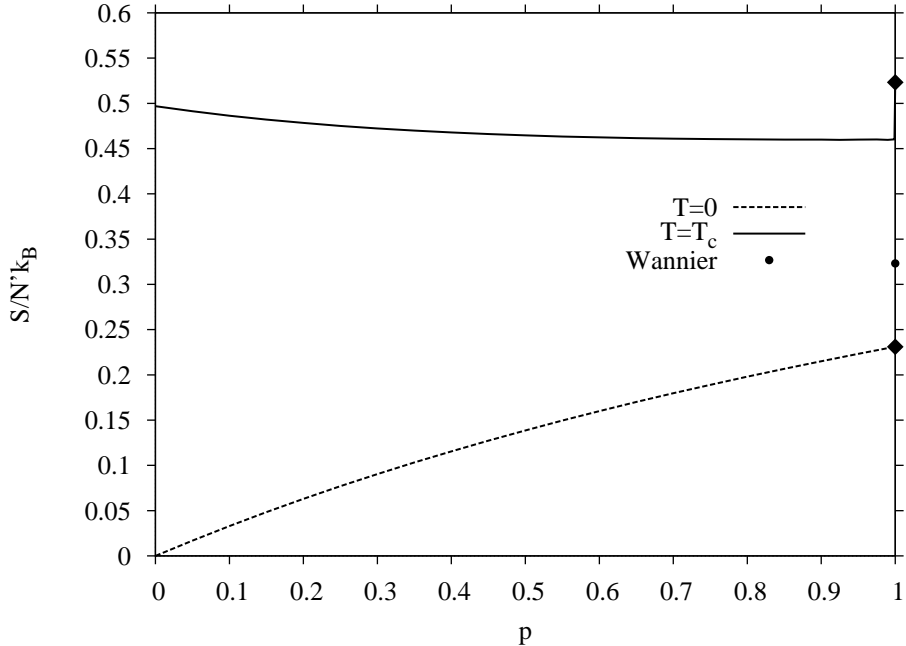}
\caption{Entropy $S/N'k_{\rm B}$ per occupied lattice site, expressed in Boltzmann constant units, vs. concentration $p$. Upper curve coresponds to entropy at critical temperature $T=T_{\rm c}$, whereas lower curve presents the residual entropy at $T=0$.}
\label{fig:fig5}
\end{figure}
Taking into account the above problems, we decided to modify the PA method in order to obtain simplified but still qualitatively correct description of the model. The result for $T_c$ is shown in Fig.~\ref{fig:fig2} as a solid curve. For $p=0$ we obtained $k_{\rm B} T_{\rm c}/|J|=1/\left(2\ln3\right)\approx$0.4551, i.e., the same value as for ferromagnetic case \cite{Balcerzak6}, and identical with Bethe result $k_{\rm B} T_{\rm c}/|J|=1/\left[2\ln\left(z/\left(z-2\right)\right)\right]$ for NN number $z=3$. On the other hand, for $p=1$ the exact Wannier result $T_c=0$ is recovered. It follows from the present method that the ordered phase exists in the full range of $0 \le p \le 1$, whereas for $p=1$ and $T=0$ it coexists with the disordered phase.\\
\begin{figure}
\includegraphics[scale=1.0]{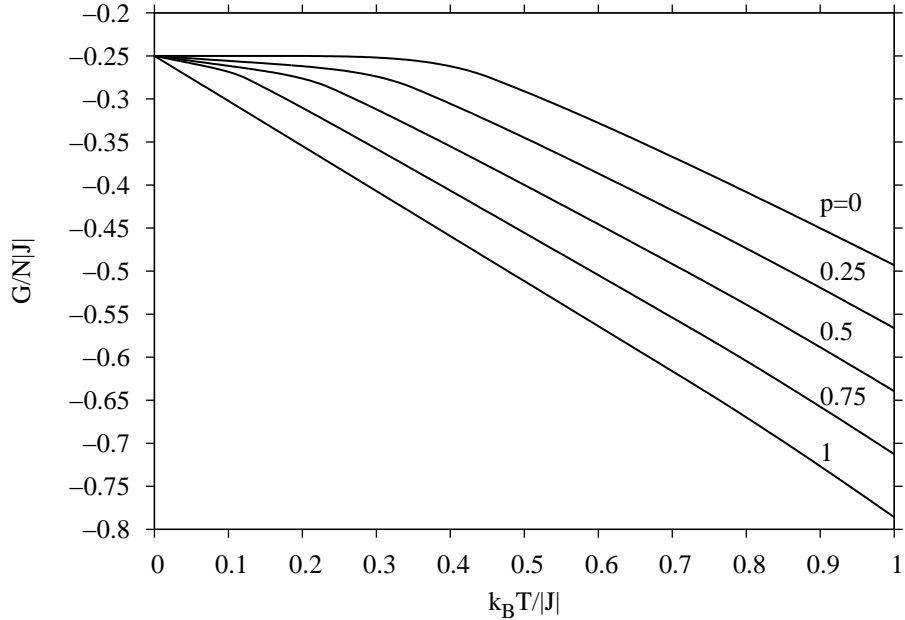}
\caption{Gibbs energy $G/N|J|$ per lattice site, expressed in $|J|$ units, vs. dimensionless temperature $k_{\rm B} T/|J|$. Different curves correspond to various concentrations $p$.}
\label{fig:fig6}
\end{figure}
In Fig.~\ref{fig:fig3} the sublattice spontaneous magnetizations $m_A=-m_B$ are shown vs. temperature. Magnetization of the diluted sublattice $C$ amounts to $pm_C=0$ for all temperatures. Different curves correspond to various concentrations $p$. For $T=0$ the sublattice magnetization is constant vs. $p$ and reaches its saturated value $m_A=1/2$ ($m_B=-1/2$), which is in agreement with the ground state. This result differs from that obtained by EFT \cite {Zukovic2}, where the magnetization at $T=0$ depends on concentration. For $p \to 1$ a jump of magnetization from the value 1/2 to 0 signals the 1st order phase transition.\\
\begin{figure}
\includegraphics[scale=1.0]{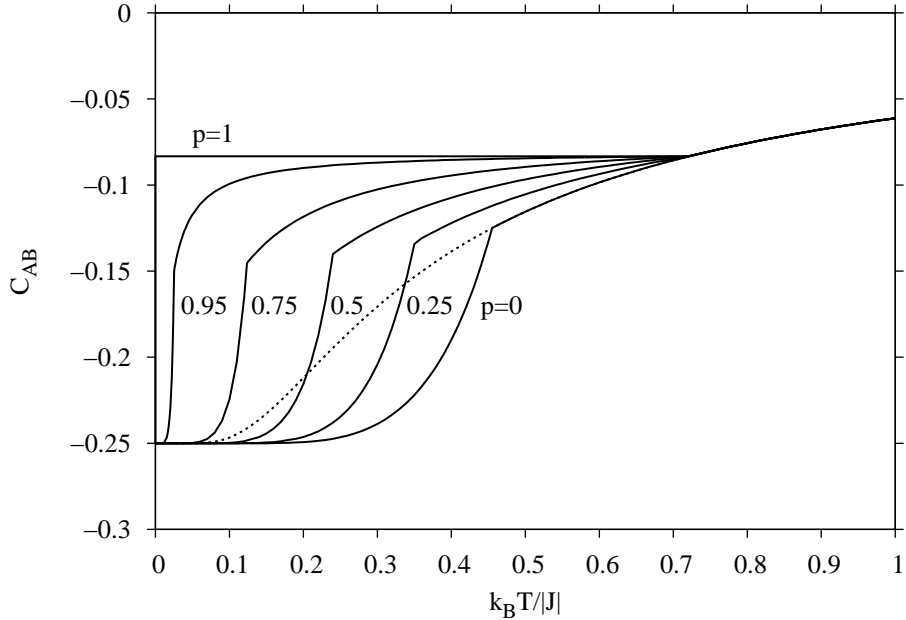}
\caption{Nearest-neighbours correlation function $c_{AB}$ vs. dimensionless temperature $k_{\rm B} T/|J|$. Different curves correspond to various concentrations $p$.}
\label{fig:fig7}
\end{figure}
Entropy vs. temperature per occupied lattice site, expressed in Boltzmann constant units, is illustrated in Fig.~\ref{fig:fig4}. The number of occupied lattice sites (spins) is denoted by $N'$, where $N'=N\left(2+p\right)/3$, and $N$ is the total number of lattice sites. Such normalization of the entropy allows to control its high-temperature limit, which for $T \to \infty$ amounts to $\ln 2$. As before, various curves correspond to different concentrations $p$.  For $p=0$ entropy amounts to 0 in the ground (fully ordered, antiferromagnetic) state. On the other hand, for $p=1$ and $T=0$ we obtained two values of the residual entropy: $S/N'k_{\rm B}=\left(\ln2\right) /3\approx 0.2310$ and $S/N'k_{\rm B}=0.5232$. These values  are marked by the bold points. A jump of the entropy at ($p=1$, $T=0$) signifies the 1st order transition between ordered and disordered phases which have been identified in the ground state.\\

We also found that for $p=1$, below the characteristic temperature $k_{\rm B} T_{\rm f}/|J|=0.721$, the entropy practically does not depend on temperature. It is connected with the fact that in this temperature range the correlation $c'$ takes the constant value $c'=-1/4$, which is at the edge of its physical region. Namely, in this case the absolute minimum of the Gibbs energy lies outside the physical region, i.e., for $c' <-1/4$, and therefore Eq.~(\ref{eq:eq33}) cannot be applied. However, when we restrict the domain of $c'$ to the physical region, i.e., $|c'|\le 1/4$, the minimal value of Gibbs energy in this domain exists at $c'=-1/4$. The situation changes for $k_{\rm B} T/|J|>0.721$ where the absolute minimum of $G$ falls into the physical region $|c'|\le 1/4$ and Eq.~(\ref{eq:eq33}) becomes effective for determination of $c'$.\\ 

We conclude that the kink on the entropy curve at $k_{\rm B} T_{\rm f}/|J|=0.721$ originates from cutting off the unphysical solution for the correlation function $c'$. It is seen for the entropy because these quantities are interrelated via minimum condition  for the Gibbs energy. In Fig.~\ref{fig:fig4} we also present the unphysical solution for the entropy curve for $p=1$ (dashed line), which results from the unmodified PA method. This entropy becomes negative for $k_{\rm B} T/|J|< \; \approx 0.321$, and reaches its mimimum value $S/N'k_{\rm B} \approx -1.386$ for $T=0$. However, the negative part of this entropy curve is not presented in the figure.
The effect of kink diminishes when $p$ decreases and for $p=0$, when the system has no frustration, it does not occur at all. Other kinks on the entropy curves for $p<1$ which occur in lower temperatures are connected with the 2nd order phase transitions from antiferromagnetic to paramagnetic phase.\\
\begin{figure}
\includegraphics[scale=1.0]{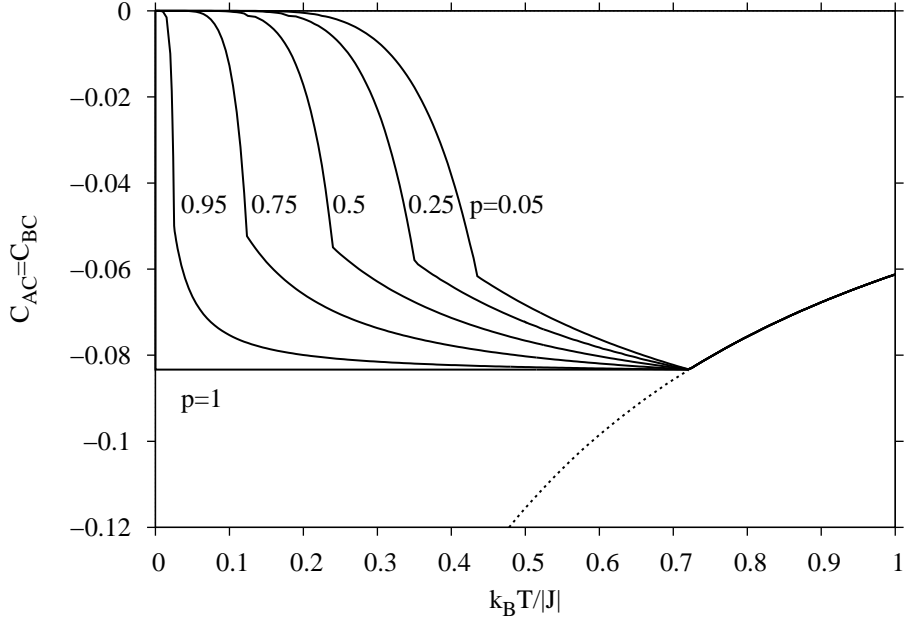}
\caption{Nearest-neighbours correlation functions $c_{AC}=c_{BC}$ vs. dimensionless temperature $k_{\rm B} T/|J|$. Different curves correspond to various concentrations $p$.}
\label{fig:fig8}
\end{figure}
Further results obtained for the entropy are illustrated in Fig.~\ref{fig:fig5} vs. concentration $p$. The upper curve presents entropy at the phase transition (N\'eel) temperature, whereas the lower curve presents entropy at $T=0$. An increasing character of the residual entropy vs. $p$ is worth noticing. For $p=1$ two values of entropy (the same as those indicated in Fig.~\ref{fig:fig4}) are seen. Since for $p=1$ we have $T_{\rm c}=0$, the entropy jump at this temperature point confirms the existence of 1st order phase transitions.  The exact Wannier result \cite{Wannier3}, i.e., $S/N'k_{\rm B}=0.32306$, is also depicted in the interval between our two points. It is worth noticing that recent result obtained from MC simulations gave the value $S/N'k_{\rm B}=0.32303$ \cite{Zukovic4}, which is very close to the exact value.\\
\begin{figure}
\includegraphics[scale=1.0]{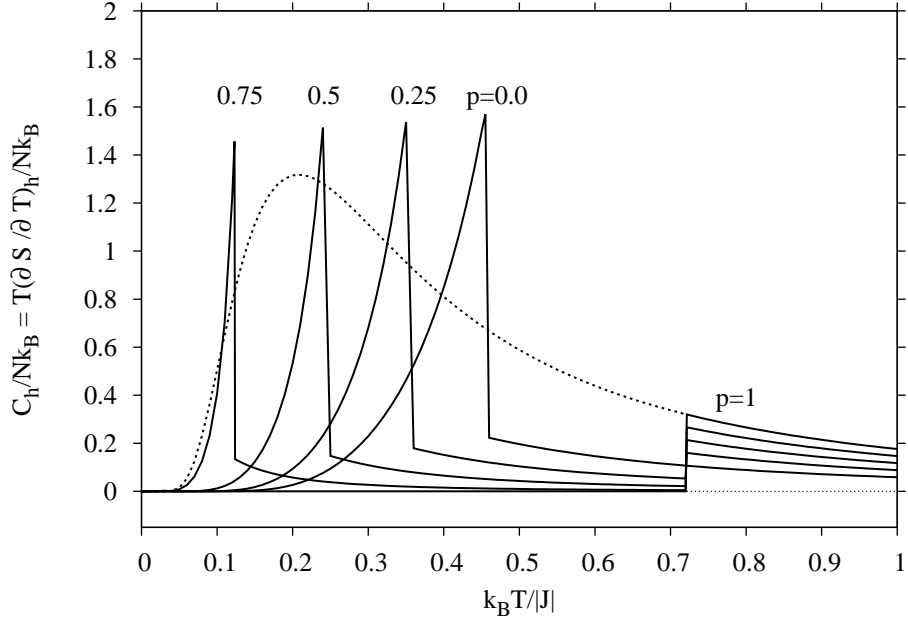}
\caption{Magnetic specific heat $C_{h}/Nk_{\rm B}$ per lattice site, expressed in Boltzmann constant units, vs. dimensionless temperature $k_{\rm B} T/|J|$. Different curves correspond to various concentrations $p$.}
\label{fig:fig9}
\end{figure}
The Gibbs energy curves vs. temperature are presented in Fig.~\ref{fig:fig6} for various concentrations $p$. A monotonously decreasing character of these curves evidences that the entropy (given by Eq.~(\ref{eq:eq30})) is positive everywhere. For large temperatures the Gibbs energy becomes linear vs. $T$ with the same slope for all curves, which corresponds to the saturation value of the entropy, as indicated in Fig.~\ref{fig:fig4}. The Gibbs energy is a smooth function vs. temperature, without any kinks for $T>0$ which would signal the 1st order phase transition. At $T=0$, for all concentrations the Gibbs energy is the same.\\
\begin{figure}
\includegraphics[scale=1.0]{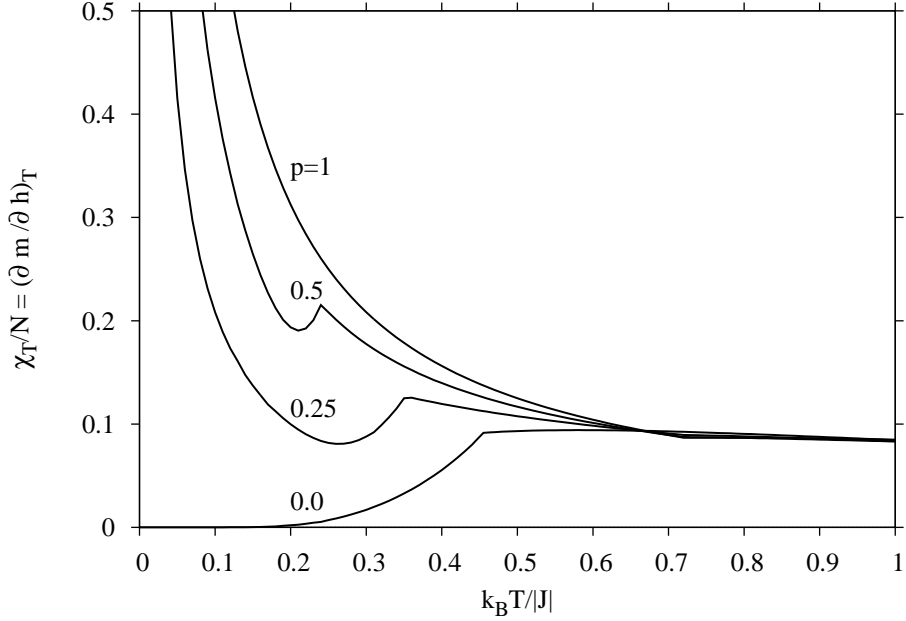}
\caption{Magnetic isothermal susceptibility $\chi_{T}/N$ per lattice site, vs. dimensionless temperature $k_{\rm B} T/|J|$. Different curves correspond to various concentrations $p$.}
\label{fig:fig10}
\end{figure}
In Fig.~\ref{fig:fig7} the NN correlation functions $c_{AB}$ are presented vs. temperature for various concentrations $p$. For $p=1$, similarly to entropy, the correlations are constant below the characteristic temperature $k_{\rm B} T_{\rm f}/|J|=0.721$, and for $T=0$ a jump of correlation function is seen between -1/4 value (for ordered phase) and -1/12 value for the disordered state. The analogous jump is seen in Fig.~\ref{fig:fig8}, where the NN corelation functions $c_{AC}=c_{BC}$ are presented. In this case a jump from the value -1/12 (for disordered phase) to 0 (for ordered state) takes place. It can be noted from Figs.~\ref{fig:fig7} and ~\ref{fig:fig8} that for $T=0$ the mean correlation per pair, $\left( c_{AB}+ c_{AC}+ c_{BC} \right)/3$, is equal to -1/12 for both states, and its absolute value amounts to one-third of the value for ferromagnetic case. The same result has also been pointed out by Wannier \cite{Wannier} in his exact solution.\\

In Figs.~\ref{fig:fig7} and \ref{fig:fig8} by dashed curves we denote the unphysical solutions for the correlation functions for $p=1$. Both curves tend to -1/4 value for $T=0$, however, in Fig.~\ref{fig:fig8} only a part of the curve is shown. These curves result from the unmodified PA method and correspond to the unphysical entropy (presented by dashed curve in Fig.~\ref{fig:fig4}). The ground state energy for such solution (for $p=1$) amounts to $-3/4|J|$ per spin and is 3 times lower than the exact ground state energy \cite{Wannier}, as well as the value obtained in the modified PA method.\\

A decreasing character of the $c_{AC}=c_{BC}$ curves in Fig.~\ref{fig:fig8}, when $T$ increases in the range $0< k_{\rm B} T/|J| \le0.721$, has no influence on the sign of magnetic specific heat, since the total internal energy, and entropy, are monotonously increasing functions of temperature. Thus, the specific heat is positive everywhere.\\

The specific heat can be conveniently calculated from Eq.~(\ref{eq:eq35}) and the results are presented in Fig.~\ref{fig:fig9}. In this figure, apart from the pronounced peaks corresponding to the N\'eel temperatures, a small jumps can be noted for $k_{\rm B} T_{\rm f}/|J|=0.721$. Again, these jumps result from the entropy (or correlation function) kinks presented in Figs.~\ref{fig:fig4},~\ref{fig:fig7} and \ref{fig:fig8}. In Fig.~\ref{fig:fig9}, by dashed curve, we present also the specific heat for $p=1$, when the correlation functions are not limited to the physical range, i.e., are calculated within the unmodified PA method. Such specific heat shows a broad maximum whose magnitude is comparable with the peaks at the phase transitions presented in this figure for $p<1$. We are aware that the broad maximum of the paramagnetic specific heat has also been found in MC simulations \cite{Robinson1}, in accordance with the exact results for triangular lattice \cite{Wannier}. However, in the modified PA method 
only a tail of this 
peak is present as a physical solution, whereas the main part has been cut off in order to obtain self-consistent thermodynamics (and to avoid unphysical entropy - see dashed line in Fig.~\ref{fig:fig4}).
It is interesting to note here that a similar double-peak structure of the magnetic specific heat has been found in the triangular antiferromagnet ${\rm Ni} \, {\rm Ga}_2 \, {\rm S}_4$ \cite{Nakatsuji}.\\

In the last figure (Fig.~\ref{fig:fig10}) the initial magnetic susceptibility (for $h=0$) is shown vs. temperature for various concentrations $p$. In this case only the finite peaks connected with the phase transition (N\'eel) temperature are seen.  For $p>0$ the susceptibility diverges at $T=0$, which is connected with the rapid rearrangement of the ground state from $m_A=-m_B=1/2$ and $m_C=0$ configuration (for $h=0$) to the configuration characterized by $m_A=-m_B=1/2$ and $m_C=1/2$, which occurs for $0< h/|J|< 3/2$. Moreover, a divergence of $\chi_T$ for $p=1$ and $T=0$, and lack of peak for $T>0$,  confirms the phase transition in the ground state for triangular antiferromagnet, in accordance with previous discussion. It can be noted from Figs.~\ref{fig:fig9} and ~\ref{fig:fig10} that both specific heat and susceptibility curves present a correct thermodynamic behaviour in the limits $T \to 0$ and $T \to \infty$.\\

%section 4
\section{Summary and final conclusions}
 
In this paper we modified the PA method in order to adopt it for frustrated systems. The Kaya-Berker model presents an ideal benchmark for testing the method, since the degree of frustration is controlled by the dilution parameter $p$. Moreover, the exact solutions in the limits $p=0$ and $p=1$ are known. Fig.~\ref{fig:fig2} illustrates that the results are extraordinarily sensitive to the approximate methods. The occurrence of unphysical solutions for $0<p<1$, or lack of complete thermodynamic description, are the main problems in all former analytical approaches. The MC simulations are very useful, however, in the frustrated systems they are difficult to perform in the low-temperature region.\\

The present method, although is relatively simple and based on the approximate Gibbs energy, gives physically correct description of all thermodynamic quantities in frustrated system. 
It allows to eliminate the unphysical solutions which result from the unmodified PA method.
In particular, the modified PA method gives qualitatively correct phase diagram as well as the exact energy and magnetization of the ground state in the full ($p,h$)-space (Sec. III). Important finding is that the critical temperature tends to zero when $p \to 1$. This is contrary to HSMF and EFT
methods, however, in agreement with the tendency seen from MC and exact Wannier result \cite{Wannier}. It has also been shown that the ordered phase for $h=0$, characterized by $m_A=-m_B$ and $m_C=0$, corresponds to the minimum of modified Gibbs potential, and such a phase is in agreement with MC results.\\

In conclusion, the modified PA method gives physically correct description of Kaya-Berker model. The results are most accurate  in low temperatures ($T \le T_{\rm c}$) and, in particular, are exact in the ground state.
Taking into account the completeness of the method, the description of the model is better than obtained by any other analytical method used to date. A price for this completeness is a less accurate description for higher temperatures ($T > T_{\rm c}$), consisting in cutting off the paramagnetic maximum of specific heat and flattening of entropy and correlation functions. However, such a cutting off was necessary in order to obtain self-consistent thermodynamics in all temperatures and to eliminate the unphysical solutions. It should be noted that an example of similar radical cutting off has already been known in thermodynamics; just to mention the Maxwell's construction for van der Waals equation of state. The flattening of the entropy and correlation function for $T_{\rm c} < T < T_{\rm f}$ appears to be most spectacular for 
triangular lattice ($p=1$), since $T_{\rm c} \to 0$ and $T_{\rm c}-T_{\rm f}$ distance is the longest. This effect of flattening vanishes gradually with increase of dilution.\\

From the analysis of the above results it becomes obvious that occurrence of characteristic temperature, $k_{\rm B} T_{\rm f}/|J|=0.721$, for which such quantities as the correlation functions, or entropy, present a kink, is an artefact of the approximation. Fortunately, this effect occurs in the paramagnetic region, far above the critical temperatures, and has no destructive influence on the low-temperature behaviour (i.e., in the most interesting regime) and on the ground state where the method is most accurate. It is also worth mentioning that in the limit $T \to \infty$, where entropy saturates, we again obtain correct thermodynamic behaviour of all calculated quantities.\\

As a final remark, we hope that the presented approach can also be useful for investigations of other spin systems with geometrical frustrations.\\  

\appendix
\section{The variational equations for $h=0$}

In order to obtain the detailed form of variational equations without external field ($h=0$) one should note that spontaneously ordered phase  is characterized by: $m_A=-m_B\equiv m$ and $m_C=0$. Then, Eqs.~(\ref{eq:eq31}) reduce to $\partial G/\partial m=0$. For the sake of simplicity we introduce a short notation:
\begin{equation}
R_1\equiv \rho^{++}_{A B}=\frac{1}{4}+c_{A B}; \;\;
R_2\equiv \rho^{+-}_{A B}=\frac{1}{4}+m-c_{A B}; \;\;
R_3\equiv \rho^{-+}_{A B}=\frac{1}{4}-m-c_{A B}; \;\;
R_4\equiv \rho^{--}_{A B}=\frac{1}{4}+c_{A B},
\label{eq:eq37}
\end{equation}
and
\begin{eqnarray}
A_1&\equiv& \rho^{++}_{A C}=\rho^{--}_{B C}=\frac{1}{4}+\frac{1}{2}m+\left(\frac{1}{2}+2c_{A B}\right)c'\nonumber\\
A_2&\equiv& \rho^{+-}_{A C}=\rho^{-+}_{B C}=\frac{1}{4}+\frac{1}{2}m-\left(\frac{1}{2}+2c_{A B}\right)c'\nonumber\\
A_3&\equiv& \rho^{-+}_{A C}=\rho^{+-}_{B C}=\frac{1}{4}-\frac{1}{2}m-\left(\frac{1}{2}+2c_{A B}\right)c'\nonumber\\
A_4&\equiv& \rho^{--}_{A C}=\rho^{++}_{B C}=\frac{1}{4}-\frac{1}{2}m+\left(\frac{1}{2}+2c_{A B}\right)c'.
\label{eq:eq38}
\end{eqnarray}
With the help of the above coefficients the equilibrium condition $\partial G/\partial m=0$ takes the form of:
\begin{equation}
\ln \left( \frac{R_2}{R_3}\right)+p \ln \left( \frac{A_1 A_2}{A_3 A_4}\right)=2\left(\frac{2}{3}+p\right)\ln \left( \frac{1/2+m}{1/2-m}\right).
\label{eq:eq39}
\end{equation} 
In turn, from Eq.~(\ref{eq:eq32}) we obtain:
\begin{equation}
\frac{|J|}{k_{\rm B}T}\left(1+4pc'\right)=4pc'\ln \left( \frac{A_2 A_3}{A_1 A_4}\right)+ \ln \left( \frac{R_2 R_3}{R_1 R_4}\right),  
\label{eq:eq40}
\end{equation} 
and Eq.~(\ref{eq:eq33}) leads to the result:
\begin{equation}
\frac{|J|}{k_{\rm B}T}=\ln \left( \frac{A_2 A_3}{A_1 A_4}\right).  
\label{eq:eq41}
\end{equation} 
Eqs.~(\ref{eq:eq39}-\ref{eq:eq41}) form a set of three variational equations for $m$, $c_{A B}$ and $c'$. However, Eq.~(\ref{eq:eq41}) should be used only if $|c'|\le 1/4$. If this is not the case, according to discussion presented in the theoretical section (subsection C), we should assume $c'=-1/4=const.$  Then, only two variational equations (Eq.~(\ref{eq:eq39}) and Eq.~(\ref{eq:eq40})) are effective. It has been checked by the direct numerical calculation of Gibbs functional that such a choice of $m$, $c_{A B}$ and $c'$ minimizes the Gibbs energy in the physical range of these parameters, whereas $h=0$.

%\bibliography{text_1}

\begin{thebibliography}{46}
\expandafter\ifx\csname natexlab\endcsname\relax\def\natexlab#1{#1}\fi
\expandafter\ifx\csname bibnamefont\endcsname\relax
  \def\bibnamefont#1{#1}\fi
\expandafter\ifx\csname bibfnamefont\endcsname\relax
  \def\bibfnamefont#1{#1}\fi
\expandafter\ifx\csname citenamefont\endcsname\relax
  \def\citenamefont#1{#1}\fi
\expandafter\ifx\csname url\endcsname\relax
  \def\url#1{\texttt{#1}}\fi
\expandafter\ifx\csname urlprefix\endcsname\relax\def\urlprefix{URL }\fi
\providecommand{\bibinfo}[2]{#2}
\providecommand{\eprint}[2][]{\url{#2}}

%\bibitem[{\citenamefont{}()}]{}
%\bibinfo{author}{\bibfnamefont{}~\bibnamefont{}},
%\bibinfo{journal}{}
%\textbf{\bibinfo{volume}{}}, \bibinfo{pages}{} (\bibinfo{year}{}).

%\bibitem[{\citenamefont{}( {\natexlab{b}})}]{}
%\bibinfo{author}{\bibfnamefont{}~\bibnamefont{}}, in
%\emph{\bibinfo{booktitle}{}}, edited by
%\bibinfo{editor}{\bibfnamefont{} \bibnamefont{}}
%(\bibinfo{publisher}{},
%\bibinfo{year}{}{\natexlab{b}}), chap.~\bibinfo{chapter}{}.

\bibitem[{\citenamefont{Wannier}(1950)}]{Wannier}
\bibinfo{author}{\bibfnamefont{G.~H.}~\bibnamefont{Wannier}},
\bibinfo{journal}{Phys. Rev.}
\textbf{\bibinfo{volume}{79}}, \bibinfo{pages}{357} (\bibinfo{year}{1950}).

\bibitem[{\citenamefont{Blackman et~al.}(1981)}]{Blackman}
\bibinfo{author}{\bibfnamefont{J.~A.}~\bibnamefont{Blackman}, ~\bibfnamefont{G.}~\bibnamefont{Kemeny}}, \bibnamefont{and}
\bibinfo{author}{\bibfnamefont{J.~P.}~\bibnamefont{Straley}},
\bibinfo{journal}{J. Phys. C: Solid State Phys.}
\textbf{\bibinfo{volume}{14}}, \bibinfo{pages}{385} (\bibinfo{year}{1981}).

\bibitem[{\citenamefont{Kawamura}(1983)}]{Kawamura}
\bibinfo{author}{\bibfnamefont{H.}~\bibnamefont{Kawamura}},
\bibinfo{journal}{Prog. Theor. Phys.}
\textbf{\bibinfo{volume}{70}}, \bibinfo{pages}{697} (\bibinfo{year}{1983}).

\bibitem[{\citenamefont{Chowdhury}(1986{\natexlab{b}})}]{Chowdhury}
\bibinfo{author}{\bibfnamefont{D.}~\bibnamefont{Chowdhury}}, 
\emph{\bibinfo{booktitle}{Spin Glasses and Other Frustrated Systems}}, 
(\bibinfo{publisher}{World Scientific Publishing Co},
\bibinfo{year}{1986}{\natexlab{b}}).

\bibitem[{\citenamefont{Fisher and Hertz}(1991{\natexlab{b}})}]{Fisher}
\bibinfo{author}{\bibfnamefont{K.~H.}~\bibnamefont{Fisher}} and
\bibinfo{author}{\bibfnamefont{J.~A.}~\bibnamefont{Hertz}}, 
\emph{\bibinfo{booktitle}{Spin Glasses}}, 
(\bibinfo{publisher}{Cambridge University Press},
\bibinfo{year}{1991}{\natexlab{b}}).

\bibitem[{\citenamefont{Diep et~al.}(1991)}]{Diep1}
\bibinfo{author}{\bibfnamefont{H.~T.}~\bibnamefont{Diep}, ~\bibfnamefont{M.}~\bibnamefont{Debauche}} \bibnamefont{and}
\bibinfo{author}{\bibfnamefont{H.}~\bibnamefont{Giacomini}},
\bibinfo{journal}{Phys. Rev. B}
\textbf{\bibinfo{volume}{43}}, \bibinfo{pages}{8759} (\bibinfo{year}{1991}).

\bibitem[{\citenamefont{Netz and Berker}(1991)}]{Netz}
\bibinfo{author}{\bibfnamefont{R.~R.}~\bibnamefont{Netz}} \bibnamefont{and}
\bibinfo{author}{\bibfnamefont{A.~N.}~\bibnamefont{Berker}},
\bibinfo{journal}{Phys. Rev. Lett.}
\textbf{\bibinfo{volume}{66}}, \bibinfo{pages}{377} (\bibinfo{year}{1991}).

\bibitem[{\citenamefont{Pelizzola and Pretti}(1999)}]{Pelizzola}
\bibinfo{author}{\bibfnamefont{A.}~\bibnamefont{Pelizzola}} \bibnamefont{and}
\bibinfo{author}{\bibfnamefont{M.}~\bibnamefont{Pretti}},
\bibinfo{journal}{Phys. Rev. B}
\textbf{\bibinfo{volume}{60}}, \bibinfo{pages}{10134} (\bibinfo{year}{1999}).

\bibitem[{\citenamefont{Kaya and Berker}(2000)}]{Kaya}
\bibinfo{author}{\bibfnamefont{H.}~\bibnamefont{Kaya}} \bibnamefont{and}
\bibinfo{author}{\bibfnamefont{A.~N.}~\bibnamefont{Berker}},
\bibinfo{journal}{Phys. Rev. E}
\textbf{\bibinfo{volume}{62}}, \bibinfo{pages}{R1469} (\bibinfo{year}{2000}).

\bibitem[{\citenamefont{Garcia-Adeva and Hubner}(2001)}]{Garcia}
\bibinfo{author}{\bibfnamefont{A.~J.}~\bibnamefont{Garcia-Adeva}} \bibnamefont{and}
\bibinfo{author}{\bibfnamefont{D.~L.}~\bibnamefont{Hubner}},
\bibinfo{journal}{Phys. Rev. B}
\textbf{\bibinfo{volume}{63}}, \bibinfo{pages}{174433} (\bibinfo{year}{2001}).

\bibitem[{\citenamefont{Moessner and Sondhi}(2001)}]{Moessner}
\bibinfo{author}{\bibfnamefont{R.}~\bibnamefont{Moessner}} \bibnamefont{and}
\bibinfo{author}{\bibfnamefont{S.~L.}~\bibnamefont{Sondhi}},
\bibinfo{journal}{Phys. Rev. B}
\textbf{\bibinfo{volume}{63}}, \bibinfo{pages}{224401} (\bibinfo{year}{2001}).

%%%

\bibitem[{\citenamefont{Galam and Koseleff}(2002)}]{Galam}
\bibinfo{author}{\bibfnamefont{S.}~\bibnamefont{Galam}} \bibnamefont{and}
\bibinfo{author}{\bibfnamefont{P.~-V.}~\bibnamefont{Koseleff}},
\bibinfo{journal}{Eur. Phys. J. B}
\textbf{\bibinfo{volume}{28}}, \bibinfo{pages}{149} (\bibinfo{year}{2002}).

\bibitem[{\citenamefont{Wu and Mattis}(2003)}]{Wu}
\bibinfo{author}{\bibfnamefont{J.}~\bibnamefont{Wu}} \bibnamefont{and}
\bibinfo{author}{\bibfnamefont{D.~C.}~\bibnamefont{Mattis}},
\bibinfo{journal}{Phys. Rev. B}
\textbf{\bibinfo{volume}{67}}, \bibinfo{pages}{224414} (\bibinfo{year}{2003}).

\bibitem[{\citenamefont{Robinson}(2003{\natexlab{b}})}]{Robinson1}
\bibinfo{author}{\bibfnamefont{M.~D.}~\bibnamefont{Robinson}},
\emph{\bibinfo{booktitle}{An Information Theoretic Study of the Ising Antiferromagnet with Quenched Vacancies on a Triangular Lattice}}, Master Thesis,
(\bibinfo{publisher}{University of Maine},
\bibinfo{year}{2003}{\natexlab{b}}).

\bibitem[{\citenamefont{Diep}(2004{\natexlab{b}})}]{Diep2}
\emph{\bibinfo{booktitle}{Frustrated Spin Systems}}, edited by
\bibinfo{editor}{\bibfnamefont{H.~T.}~\bibnamefont{Diep}}, 
(\bibinfo{publisher}{World Scientific Publishing Co},
\bibinfo{year}{2004}{\natexlab{b}}).





\bibitem[{\citenamefont{Nakatsuji et al.}(2005)}]{Nakatsuji}
\bibinfo{author}{\bibfnamefont{S.} ~\bibnamefont{Nakatsuji}, ~\bibnamefont{Y.}  ~\bibnamefont{Nambu}, ~\bibnamefont{H.} ~\bibnamefont{Tonomura}, ~\bibnamefont{O.} ~\bibnamefont{Sakai}, ~\bibnamefont{S.} ~\bibnamefont{Jonas}, ~\bibnamefont{C.} ~\bibnamefont{Broholm}, ~\bibnamefont{H.} ~\bibnamefont{Tsunetsugu}, ~\bibnamefont{Y.} ~\bibnamefont{Qiu}} \bibnamefont{and}
\bibinfo{author}{\bibfnamefont{Y.}~\bibnamefont{Maeno}},
\bibinfo{journal}{Science}
\textbf{\bibinfo{volume}{309}}, \bibinfo{pages}{1697} (\bibinfo{year}{2005}).




\bibitem[{\citenamefont{Ye at al.}(2006)}]{Ye}
\bibinfo{author}{\bibfnamefont{F.}~\bibnamefont{Ye}, ~\bibnamefont{Y.} ~\bibnamefont{Ren}, ~\bibnamefont{Q.} ~\bibnamefont{Huang}, ~\bibnamefont{J.~A.} ~\bibnamefont{Fernandez-Baca}, ~\bibnamefont{P.} ~\bibnamefont{Dai}, ~\bibnamefont{J.~W.} ~\bibnamefont{Lynn}} \bibnamefont{and}
\bibinfo{author}{\bibfnamefont{T.}~\bibnamefont{Kimura}},
\bibinfo{journal}{Phys. Rev. B}
\textbf{\bibinfo{volume}{73}}, \bibinfo{pages}{320404(R)} (\bibinfo{year}{2006}).

\bibitem[{\citenamefont{Itou et al.}(2008)}]{Itou}
\bibinfo{author}{\bibfnamefont{T.}~\bibnamefont{Itou},~\bibfnamefont{A.}~\bibnamefont{Oyamada},~\bibfnamefont{S.}~\bibnamefont{Maegawa},~\bibfnamefont{M.}~\bibnamefont{Tamura}} \bibnamefont{and}
\bibinfo{author}{\bibfnamefont{R.}~\bibnamefont{Kato}},
\bibinfo{journal}{Phys. Rev. B}
\textbf{\bibinfo{volume}{77}}, \bibinfo{pages}{104413} (\bibinfo{year}{2008}).

\bibitem[{\citenamefont{Andrews et al.}(2009)}]{Andrews}
\bibinfo{author}{\bibfnamefont{S.}~\bibnamefont{Andrews},~\bibfnamefont{H.}~\bibnamefont{De Sterck},~\bibfnamefont{S.}~\bibnamefont{Inglis}} \bibnamefont{and}
\bibinfo{author}{\bibfnamefont{R.~G.}~\bibnamefont{Melko}},
\bibinfo{journal}{Phys. Rev. E}
\textbf{\bibinfo{volume}{79}}, \bibinfo{pages}{041127} (\bibinfo{year}{2009}).

\bibitem[{\citenamefont{Kalz et al.}(2009)}]{Kalz}
\bibinfo{author}{\bibfnamefont{A.}~\bibnamefont{Kalz},~\bibfnamefont{A.}~\bibnamefont{Honecker},~\bibfnamefont{S.}~\bibnamefont{Fuchs}} \bibnamefont{and}
\bibinfo{author}{\bibfnamefont{T.}~\bibnamefont{Pruschke}},
\bibinfo{journal}{J. Phys.: Conf. Series}
\textbf{\bibinfo{volume}{145}}, \bibinfo{pages}{012051} (\bibinfo{year}{2009}).

\bibitem[{\citenamefont{Lacroix et~al.}(2011{\natexlab{b}})}]{Lacroix}
\emph{\bibinfo{booktitle}{Introduction to Frustrated Magnetism}}, edited by
\bibinfo{editor}{\bibfnamefont{C.} ~\bibnamefont{Lacroix}, ~\bibfnamefont{P.} ~\bibnamefont{Mendels}, and ~\bibfnamefont{F.} ~\bibnamefont{Mila},}
(\bibinfo{publisher}{Springer-Verlag},
\bibinfo{year}{2011}{\natexlab{b}}).

\bibitem[{\citenamefont{Balents}(2010)}]{Balents}
\bibinfo{author}{\bibfnamefont{L.}~\bibnamefont{Balents}},
\bibinfo{journal}{Nature}
\textbf{\bibinfo{volume}{464}}, \bibinfo{pages}{199} (\bibinfo{year}{2010}).

\bibitem[{\citenamefont{H\"artler et al.}(2010)}]{Hartler}
\bibinfo{author}{\bibfnamefont{M.}~\bibnamefont{H\"artler},~\bibfnamefont{J.}~\bibnamefont{Richter},~\bibfnamefont{D.}~\bibnamefont{Ihle}} \bibnamefont{and}
\bibinfo{author}{\bibfnamefont{S.~-L.}~\bibnamefont{Drechsler}},
\bibinfo{journal}{Phys. Rev. B}
\textbf{\bibinfo{volume}{81}}, \bibinfo{pages}{174421} (\bibinfo{year}{2010}).

\bibitem[{\citenamefont{\v{Z}ukovi\v{c} et al.}(2010)}]{Zukovic1}
\bibinfo{author}{\bibfnamefont{M.}~\bibnamefont{\v{Z}ukovi\v{c},~\bibfnamefont{M.}~\bibnamefont{Borovsk\'{y}}}} \bibnamefont{and}
\bibinfo{author}{\bibfnamefont{A.}~\bibnamefont{Bob\'{a}k}},
\bibinfo{journal}{Phys. Lett. A}
\textbf{\bibinfo{volume}{374}}, \bibinfo{pages}{4260} (\bibinfo{year}{2010}).

\bibitem[{\citenamefont{Farnell et al.}(2011)}]{Farnell}
\bibinfo{author}{\bibfnamefont{D.~J.~J.}~\bibnamefont{Farnell},~\bibfnamefont{R.~F.}~\bibnamefont{Bishop},~\bibfnamefont{P.~H.~Y.}~\bibnamefont{Li},~\bibfnamefont{J.}~\bibnamefont{Richter}} \bibnamefont{and}
\bibinfo{author}{\bibfnamefont{C.~E.}~\bibnamefont{Campbell}},
\bibinfo{journal}{Phys. Rev. B}
\textbf{\bibinfo{volume}{84}}, \bibinfo{pages}{012403} (\bibinfo{year}{2011}).

\bibitem[{\citenamefont{Albuquerque et al.}(2011)}]{Albuquerque}
\bibinfo{author}{\bibfnamefont{A.~F.}~\bibnamefont{Albuquerque},~\bibfnamefont{D.}~\bibnamefont{Schwandt},~\bibfnamefont{B.}~\bibnamefont{Het\'enyi},~\bibfnamefont{S.}~\bibnamefont{Caponi},~\bibfnamefont{M.}~\bibnamefont{Mambrini}} \bibnamefont{and}
\bibinfo{author}{\bibfnamefont{A.~M.}~\bibnamefont{L\"auchli}},
\bibinfo{journal}{Phys. Rev. B}
\textbf{\bibinfo{volume}{84}}, \bibinfo{pages}{024406} (\bibinfo{year}{2011}).

\bibitem[{\citenamefont{Hauke et al.}(2011)}]{Hauke}
\bibinfo{author}{\bibfnamefont{P.}~\bibnamefont{Hauke},~\bibfnamefont{T.}~\bibnamefont{Roscilde},~\bibfnamefont{V.}~\bibnamefont{Murg},~\bibfnamefont{J.~I.}~\bibnamefont{Cirac}} \bibnamefont{and}
\bibinfo{author}{\bibfnamefont{R.}~\bibnamefont{Schmied}},
\bibinfo{journal}{New Journal of Phys.}
\textbf{\bibinfo{volume}{13}}, \bibinfo{pages}{075017} (\bibinfo{year}{2011}).

\bibitem[{\citenamefont{Robinson et al.}(2011)}]{Robinson2}
\bibinfo{author}{\bibfnamefont{M.~D.}~\bibnamefont{Robinson},~\bibfnamefont{D.~P.}~\bibnamefont{Feldman}} \bibnamefont{and}
\bibinfo{author}{\bibfnamefont{S.~R.}~\bibnamefont{McKay}},
\bibinfo{journal}{Chaos}
\textbf{\bibinfo{volume}{21}}, \bibinfo{pages}{037114} (\bibinfo{year}{2011}).

\bibitem[{\citenamefont{Mezzacapo and Boninsegni}(2011)}]{Mezzacapo}
\bibinfo{author}{\bibfnamefont{F.}~\bibnamefont{Mezzacapo}} \bibnamefont{and}
\bibinfo{author}{\bibfnamefont{M.}~\bibnamefont{Boninsegni}},
\bibinfo{journal}{Phys. Rev. B}
\textbf{\bibinfo{volume}{85}}, \bibinfo{pages}{060402(R)} (\bibinfo{year}{2012}).

\bibitem[{\citenamefont{Fishman}(2011)}]{Fishman}
\bibinfo{author}{\bibfnamefont{R.~S.}~\bibnamefont{Fishman}},
\bibinfo{journal}{Phys. Rev. Lett.}
\textbf{\bibinfo{volume}{106}}, \bibinfo{pages}{037206} (\bibinfo{year}{2011}).

\bibitem[{\citenamefont{\v{Z}ukovi\v{c} et al.}(2012)}]{Zukovic2}
\bibinfo{author}{\bibfnamefont{M.}~\bibnamefont{\v{Z}ukovi\v{c},~\bibfnamefont{M.}~\bibnamefont{Borovsk\'{y}}}} \bibnamefont{and}
\bibinfo{author}{\bibfnamefont{A.}~\bibnamefont{Bob\'{a}k}},
\bibinfo{journal}{J. Magn. Magn. Mater.}
\textbf{\bibinfo{volume}{324}}, \bibinfo{pages}{2687} (\bibinfo{year}{2012}).

\bibitem[{\citenamefont{Shirata et al.}(2012)}]{Shirata}
\bibinfo{author}{\bibfnamefont{Y.}~\bibnamefont{Shirata},~\bibfnamefont{H.}~\bibnamefont{Tanaka},~\bibfnamefont{A.}~\bibnamefont{Matsuo}} \bibnamefont{and}
\bibinfo{author}{\bibfnamefont{K.}~\bibnamefont{Kindo}},
\bibinfo{journal}{Phys. Rev. Lett.}
\textbf{\bibinfo{volume}{108}}, \bibinfo{pages}{057205} (\bibinfo{year}{2012}).

\bibitem[{\citenamefont{Rojas et al.}(2012)}]{Rojas}
\bibinfo{author}{\bibfnamefont{M.}~\bibnamefont{Rojas},~\bibfnamefont{O.}~\bibnamefont{Rojas}} \bibnamefont{and}
\bibinfo{author}{\bibfnamefont{S.~M.}~\bibnamefont{de Souza}},
\bibinfo{journal}{Phys. Rev. E}
\textbf{\bibinfo{volume}{86}}, \bibinfo{pages}{051116} (\bibinfo{year}{2012}).

\bibitem[{\citenamefont{Dublenych}(2012)}]{Dublenych}
\bibinfo{author}{\bibfnamefont{Yu.~I.}~\bibnamefont{Dublenych}},
\bibinfo{journal}{Phys. Rev. Lett.}
\textbf{\bibinfo{volume}{109}}, \bibinfo{pages}{167202} (\bibinfo{year}{2012}).



\bibitem[{\citenamefont{\v{Z}ukovi\v{c} and Bob\'{a}k}(2013)}]{Zukovic3}
\bibinfo{author}{\bibfnamefont{M.}~\bibnamefont{\v{Z}ukovi\v{c}}} \bibnamefont{and}
\bibinfo{author}{\bibfnamefont{A.}~\bibnamefont{Bob\'{a}k}},
\bibinfo{journal}{Phys. Rev. E}
\textbf{\bibinfo{volume}{87}}, \bibinfo{pages}{032121} (\bibinfo{year}{2013}).

\bibitem[{\citenamefont{Iglovikov et al.}(2013)}]{Iglovikov}
\bibinfo{author}{\bibfnamefont{V.~I.}~\bibnamefont{Iglovikov}, ~\bibfnamefont{R.~I.}~\bibnamefont{Scalettar}} \bibnamefont{and}
\bibinfo{author}{\bibfnamefont{R.~R.~P.}~\bibnamefont{Singh}},
\bibinfo{journal}{Phys. Rev. B}
\textbf{\bibinfo{volume}{87}}, \bibinfo{pages}{214415} (\bibinfo{year}{2013}).

\bibitem[{\citenamefont{Maryasin and Zhitomirsky}(2013)}]{Maryasin}
\bibinfo{author}{\bibfnamefont{V.~S.}~\bibnamefont{Maryasin}} \bibnamefont{and}
\bibinfo{author}{\bibfnamefont{M.~E.}~\bibnamefont{Zhitomirsky}},
\bibinfo{journal}{Phys. Rev. Lett.}
\textbf{\bibinfo{volume}{111}}, \bibinfo{pages}{247201} (\bibinfo{year}{2013}).

\bibitem[{\citenamefont{Nakayama et al.}(2013)}]{Nakayama}
\bibinfo{author}{\bibfnamefont{G.}~\bibnamefont{Nakayama}, ~\bibfnamefont{S.}~\bibnamefont{Hara}, ~\bibfnamefont{H.}~\bibnamefont{Sato}, ~\bibfnamefont{Y.}~\bibnamefont{Narumi}} \bibnamefont{and}
\bibinfo{author}{\bibfnamefont{H.}~\bibnamefont{Nojiri}},
\bibinfo{journal}{J. Phys.: Condens. Matter}
\textbf{\bibinfo{volume}{25}}, \bibinfo{pages}{116003} (\bibinfo{year}{2013}).



\bibitem[{\citenamefont{Melchert and Hartmann}(2013)}]{Melchert}
\bibinfo{author}{\bibfnamefont{O.}~\bibnamefont{Melchert}} \bibnamefont{and}
\bibinfo{author}{\bibfnamefont{A.~K.}~\bibnamefont{Hartmann}},
\bibinfo{journal}{Phys. Rev. E}
\textbf{\bibinfo{volume}{87}}, \bibinfo{pages}{022107} (\bibinfo{year}{2013}).

\bibitem[{\citenamefont{Kulagin et al.}(2013)}]{Kulagin}
\bibinfo{author}{\bibfnamefont{S.~A.}~\bibnamefont{Kulagin}, ~\bibfnamefont{N.}~\bibnamefont{Prokof'ev}, ~\bibfnamefont{O.~A.}~\bibnamefont{Starykh}, ~\bibfnamefont{B.}~\bibnamefont{Svistunov}} \bibnamefont{and}
\bibinfo{author}{\bibfnamefont{C.~N.}~\bibnamefont{Varney}},
\bibinfo{journal}{Phys. Rev. B}
\textbf{\bibinfo{volume}{87}}, \bibinfo{pages}{024407} (\bibinfo{year}{2013}).


\bibitem[{\citenamefont{Yokota}(2014)}]{Yokota}
\bibinfo{author}{\bibfnamefont{T.}~\bibnamefont{Yokota}},
\bibinfo{journal}{Phys. Rev. E}
\textbf{\bibinfo{volume}{89}}, \bibinfo{pages}{012128} (\bibinfo{year}{2014}).







\bibitem[{\citenamefont{Collins and Petrenko}(1997)}]{Collins}
\bibinfo{author}{\bibfnamefont{M.~F.}~\bibnamefont{Collins}} \bibnamefont{and}
\bibinfo{author}{\bibfnamefont{O.~A.}~\bibnamefont{Petrenko}},
\bibinfo{journal}{Can. J. Phys.}
\textbf{\bibinfo{volume}{75}}, \bibinfo{pages}{605} (\bibinfo{year}{1997}).


\bibitem[{\citenamefont{Nakatsuji et al.}(2010)}]{Nakatsuji2}
\bibinfo{author}{\bibfnamefont{S.} ~\bibnamefont{Nakatsuji}, ~\bibnamefont{Y.}  ~\bibnamefont{Nambu}, \bibnamefont{and} \bibfnamefont{S.}~\bibnamefont{Onoda}},
\bibinfo{journal}{J. Phys. Soc. Jpn.}
\textbf{\bibinfo{volume}{79}}, \bibinfo{pages}{011003} (\bibinfo{year}{2010}).


\bibitem[{\citenamefont{Myoung et al.}(2011)}]{Myoung}
\bibinfo{author}{\bibfnamefont{B.~R.} ~\bibnamefont{Myoung}, ~\bibnamefont{C.~M.}  ~\bibnamefont{Kim}, ~\bibnamefont{S.~J.}  ~\bibnamefont{Kim}, ~\bibnamefont{T.}  ~\bibnamefont{Kouh}, ~\bibnamefont{Y.}  ~\bibnamefont{Hirose}, ~\bibnamefont{T.}  ~\bibnamefont{Hasegawa}, \bibnamefont{and} \bibfnamefont{C.~S.}~\bibnamefont{Kim}},
\bibinfo{journal}{J. Appl. Phys.}
\textbf{\bibinfo{volume}{109}}, \bibinfo{pages}{07E133} (\bibinfo{year}{2011}).


\bibitem[{\citenamefont{Cava et al.}(2011)}]{Cava}
\bibinfo{author}{\bibfnamefont{R.~J.}~\bibnamefont{Cava}, \bibfnamefont{K.~L.}~\bibnamefont{Holman}}, \bibfnamefont{T.}~\bibnamefont{McQueen}, \bibfnamefont{E.~J.}~\bibnamefont{Welsh}, \bibfnamefont{D.~V.}~\bibnamefont{Vest} and \bibfnamefont{A.~J.}~\bibnamefont{Williams}, in
\emph{\bibinfo{booktitle}{Introduction to Frustrated Magnetism: Materials, Experiments, Theory}}, edited by
\bibinfo{editor}{\bibfnamefont{C.} ~\bibnamefont{Lacroix}, \bibfnamefont{P.} ~\bibnamefont{Mendels} and ~\bibfnamefont{F.} ~\bibnamefont{Mila}}, Springer Series in Solid State Sciences, Vol. 164,  
(\bibinfo{publisher}{Springer, Berlin, Heidelberg},
\bibinfo{year}{2011}), p. \bibinfo{pages}{131}.

\bibitem[{\citenamefont{Loh et al.}(2008)}]{Loh}
\bibinfo{author}{\bibfnamefont{Y.~L.}~\bibnamefont{Loh}}, \bibinfo{author}{\bibfnamefont{D.~X.}~\bibnamefont{Yao}} \bibnamefont{and}
\bibinfo{author}{\bibfnamefont{E.~W.}~\bibnamefont{Carlson}},
\bibinfo{journal}{Phys. Rev. B}
\textbf{\bibinfo{volume}{77}}, \bibinfo{pages}{134402} (\bibinfo{year}{2008}).


\bibitem[{\citenamefont{Strecka}(2013)}]{Strecka}
\bibinfo{author}{\bibfnamefont{J.}~\bibnamefont{\v{C}is\'{a}rov\'{a}}} \bibnamefont{and}
\bibinfo{author}{\bibfnamefont{J.}~\bibnamefont{Stre\v{c}ka}},
\bibinfo{journal}{Phys. Rev. B}
\textbf{\bibinfo{volume}{87}}, \bibinfo{pages}{024421} (\bibinfo{year}{2013}).


\bibitem[{\citenamefont{Zhang et al.}(1994)}]{Zhang}
\bibinfo{author}{\bibfnamefont{G.~M.}~\bibnamefont{Zhang}, ~\bibfnamefont{C.~Z.}~\bibnamefont{Yang}}, 
\bibinfo{journal}{Phys. Rev. B}
\textbf{\bibinfo{volume}{50}}, \bibinfo{pages}{12546} (\bibinfo{year}{1994}).

\bibitem[{\citenamefont{Jacobsen et al.}(1997)}]{Jacobsen}
\bibinfo{author}{\bibfnamefont{J.~L.}~\bibnamefont{Jacobsen}, ~\bibfnamefont{H.~C.}~\bibnamefont{Fogedby}}, 
\bibinfo{journal}{Physica A}
\textbf{\bibinfo{volume}{246}}, \bibinfo{pages}{563} (\bibinfo{year}{1997}).


\bibitem[{\citenamefont{Stephenson}(1964)}]{Stephenson}
\bibinfo{author}{\bibfnamefont{J.}~\bibnamefont{Stephenson}},
\bibinfo{journal}{J. Math. Phys.}
\textbf{\bibinfo{volume}{5}}, \bibinfo{pages}{1009} (\bibinfo{year}{1964}).

\bibitem[{\citenamefont{Balcerzak}(2008)\citenamefont{Balcerzak}}]{Balcerzak1}
\bibinfo{author}{\bibfnamefont{T.}~\bibnamefont{Balcerzak}},
\bibinfo{journal}{J. Magn. Magn. Mater.} \textbf{\bibinfo{volume}{320}},
\bibinfo{pages}{2359} (\bibinfo{year}{2008}).  

\bibitem[{\citenamefont{Honmura}(1979)\citenamefont{Honmura, Kaneyoshi}}]{Honmura}
\bibinfo{author}{\bibfnamefont{R.}~\bibnamefont{Honmura}},
\bibinfo{author}{\bibfnamefont{T.}~\bibnamefont{Kaneyoshi}},
\bibinfo{journal}{J. Phys. C} \textbf{\bibinfo{volume}{12}},
\bibinfo{pages}{3979} (\bibinfo{year}{1979}).   

\bibitem[{\citenamefont{Matsudaira}(1973)\citenamefont{Matsudaira}}]{Matsudaira}
\bibinfo{author}{\bibfnamefont{N.}~\bibnamefont{Matsudaira}},
\bibinfo{journal}{J. Phys. Soc. Japan} \textbf{\bibinfo{volume}{35}},
\bibinfo{pages}{1593} (\bibinfo{year}{1973}).  

\bibitem[{\citenamefont{Balcerzak}(1985)\citenamefont{Balcerzak, Bobak, Mielnicki, Truong}}]{Balcerzak2}
\bibinfo{author}{\bibfnamefont{T.}~\bibnamefont{Balcerzak}},
\bibinfo{author}{\bibfnamefont{A.}~\bibnamefont{Bob\'ak}},
\bibinfo{author}{\bibfnamefont{J.}~\bibnamefont{Mielnicki}},
\bibinfo{author}{\bibfnamefont{V. H.}~\bibnamefont{Truong}},
\bibinfo{journal}{Phys. Stat. Sol. (b)} \textbf{\bibinfo{volume}{130}},
\bibinfo{pages}{183} (\bibinfo{year}{1985}). 

\bibitem[{\citenamefont{Banavar}(1991)\citenamefont{Banavar, Cieplak, Maritan}}]{Banavar}
\bibinfo{author}{\bibfnamefont{J. N.}~\bibnamefont{Banavar}},
\bibinfo{author}{\bibfnamefont{M.}~\bibnamefont{Cieplak}},
\bibinfo{author}{\bibfnamefont{A.}~\bibnamefont{Maritan}},
\bibinfo{journal}{Phys. Rev. Lett.} \textbf{\bibinfo{volume}{67}},
\bibinfo{pages}{1807} (\bibinfo{year}{1991}).  

\bibitem[{\citenamefont{Parisi}(1988)\citenamefont{Parisi}}]{Parisi}
\bibinfo{author}{\bibfnamefont{G.}~\bibnamefont{Parisi}},
\emph{\bibinfo{booktitle}{Statistical Field Theory}}
(\bibinfo{publisher}{Addison-Wesley, Reading, MA}, \bibinfo{year}{1988}).

\bibitem[{\citenamefont{Bukman et al.}(1991)}]{Bukman}
\bibinfo{author}{\bibfnamefont{D.~J.}~\bibnamefont{Bukman}, ~\bibfnamefont{G.}~\bibnamefont{An}} \bibnamefont{and}
\bibinfo{author}{\bibfnamefont{J.~M.~J.}~\bibnamefont{van Leeuven}},
\bibinfo{journal}{Phys. Rev. B}
\textbf{\bibinfo{volume}{43}}, \bibinfo{pages}{13352} (\bibinfo{year}{1991}).

\bibitem[{\citenamefont{Balcerzak}(2003)}]{Balcerzak3}
\bibinfo{author}{\bibfnamefont{T.}~\bibnamefont{Balcerzak}},
\bibinfo{journal}{Physica A}
\textbf{\bibinfo{volume}{317}}, \bibinfo{pages}{213} (\bibinfo{year}{2003}).

\bibitem[{\citenamefont{Balcerzak and \L u\.zniak}(2009)}]{Balcerzak4}
\bibinfo{author}{\bibfnamefont{T.}~\bibnamefont{Balcerzak}} \bibnamefont{and}
\bibinfo{author}{\bibfnamefont{I.}~\bibnamefont{}\L u\.zniak},
\bibinfo{journal}{Physica A}
\textbf{\bibinfo{volume}{388}}, \bibinfo{pages}{357} (\bibinfo{year}{2009}).

\bibitem[{\citenamefont{Balcerzak and Sza{\l}owski}(2009)}]{Balcerzak5}
\bibinfo{author}{\bibfnamefont{T.}~\bibnamefont{Balcerzak}} \bibnamefont{and}
\bibinfo{author}{\bibfnamefont{K.}~\bibnamefont{Sza{\l}owski}},
\bibinfo{journal}{Phys. Rev. B}
\textbf{\bibinfo{volume}{80}}, \bibinfo{pages}{144404} (\bibinfo{year}{2009}).

\bibitem[{\citenamefont{Balcerzak et al.}(2012)}]{Balcerzak6}
\bibinfo{author}{\bibfnamefont{T.}~\bibnamefont{Balcerzak}, ~\bibfnamefont{K.}~\bibnamefont{Sza{\l}owski}, ~\bibfnamefont{M.}~\bibnamefont{\v{Z}ukovi\v{c}}, ~\bibfnamefont{M.}~\bibnamefont{Borovsk\'{y}}, ~\bibfnamefont{A.}~\bibnamefont{Bob\'{a}k}} \bibnamefont{and}
\bibinfo{author}{\bibfnamefont{M.}~\bibnamefont{Ja\v{s}\v{c}ur}}, 
\bibinfo{journal}{Acta Physicae Superficierum}~(\bibinfo{publisher}{University of \L\'{o}d\'{z}}),
\textbf{\bibinfo{volume}{XII}}, \bibinfo{pages}{34} (\bibinfo{year}{2012}); 
\bibinfo{journal}{arXiv: 1211.1283v2}
(\bibinfo{year}{2013}). 

\bibitem[{\citenamefont{Schick and Walker}(1977)}]{Schick}
\bibinfo{author}{\bibfnamefont{M.}~\bibnamefont{Schick}} \bibnamefont{and}
\bibinfo{author}{\bibfnamefont{J.~S.}~\bibnamefont{Walker}},
\bibinfo{journal}{Phys. Rev. B}
\textbf{\bibinfo{volume}{16}}, \bibinfo{pages}{2205} (\bibinfo{year}{1977}).

\bibitem[{\citenamefont{Wannier}(1945)}]{Wannier2}
\bibinfo{author}{\bibfnamefont{G.~H.}~\bibnamefont{Wannier}},
\bibinfo{journal}{Revs. Mod. Phys.}
\textbf{\bibinfo{volume}{17}}, \bibinfo{pages}{50} (\bibinfo{year}{1945}).

\bibitem[{\citenamefont{Domb}(1972 {\natexlab{b}})}]{Domb}
\bibinfo{author}{\bibfnamefont{C.}~\bibnamefont{Domb}}, in
\emph{\bibinfo{booktitle}{Phase Transitions and Critical Phenomena}}, Vol. 3, chap.~\bibinfo{chapter}{6}, edited by
\bibinfo{editor}{\bibfnamefont{C.} ~\bibnamefont{Domb} and ~\bibfnamefont{M.~S.} ~\bibnamefont{Green}} 
(\bibinfo{publisher}{Academic Press},
\bibinfo{year}{1972}{\natexlab{b}}).

\bibitem[{\citenamefont{Dresselhaus}(1965)}]{Dresselhaus}
\bibinfo{author}{\bibfnamefont{G.}~\bibnamefont{Dresselhaus}},
\bibinfo{journal}{Phys. Rev.}
\textbf{\bibinfo{volume}{139}}, \bibinfo{pages}{A855} (\bibinfo{year}{1965}).

\bibitem[{\citenamefont{Yamamoto}(2009)}]{Yamamoto}
\bibinfo{author}{\bibfnamefont{D.}~\bibnamefont{Yamamoto}},
\bibinfo{journal}{Phys. Rev. B}
\textbf{\bibinfo{volume}{79}}, \bibinfo{pages}{144427} (\bibinfo{year}{2009}).

\bibitem[{\citenamefont{Crest and Gabl}(1977)}]{Crest}
\bibinfo{author}{\bibfnamefont{G.~S.}~\bibnamefont{Crest}} \bibnamefont{and}
\bibinfo{author}{\bibfnamefont{E.~G.}~\bibnamefont{Gabl}},
\bibinfo{journal}{Phys. Rev. Lett.}
\textbf{\bibinfo{volume}{43}}, \bibinfo{pages}{1182} (\bibinfo{year}{1979}).

\bibitem[{\citenamefont{Katsura}(1996 {\natexlab{b}})}]{Katsura}
\bibinfo{author}{\bibfnamefont{S.}~\bibnamefont{Katsura}}, in
\emph{\bibinfo{booktitle}{Theory and Applications of the Cluster Variation and Path Probability Methods}}, edited by
\bibinfo{editor}{\bibfnamefont{J.~L.} ~\bibnamefont{Mor\'an Lop\'ez} and ~\bibfnamefont{J.~M.} ~\bibnamefont{Sanchez}} 
(\bibinfo{publisher}{Plenum Press},
\bibinfo{year}{1996}{\natexlab{b}}).

\bibitem[{\citenamefont{Wannier}(1973)}]{Wannier3}
\bibinfo{author}{\bibfnamefont{G.~H.}~\bibnamefont{Wannier}},
\bibinfo{journal}{Phys. Rev. B}
\textbf{\bibinfo{volume}{7}}, \bibinfo{pages}{5017} (\bibinfo{year}{1973}).

\bibitem[{\citenamefont{\v{Z}ukovi\v{c}}(2013)}]{Zukovic4}
\bibinfo{author}{\bibfnamefont{M.}~\bibnamefont{\v{Z}ukovi\v{c}}},
\bibinfo{journal}{Eur. Phys. Journal B}
\textbf{\bibinfo{volume}{86}}, \bibinfo{pages}{283} (\bibinfo{year}{2013}).

%%%%%%%%%%
\end{thebibliography}

\end{document}